\documentclass[aps,prd,amsmath,amssymb,footinbib]{revtex4-2}

\usepackage{graphicx}
\usepackage{natbib}

\usepackage{bm}
\usepackage{url}
\usepackage{textcase}
\usepackage{subcaption}
\usepackage{float}
\usepackage{amsthm}
\usepackage{braket}
\usepackage{color}
\usepackage{xcolor}
\usepackage[normalem]{ulem}

\newcommand{\be}{\begin{equation}}
\newcommand{\ee}{\end{equation}}

\begin{document}

\title{Equatorial extreme-mass-ratio inspirals in Kerr black holes with scalar hair spacetimes}

\author{Lucas G. Collodel}
 \email{lucas.gardai-collodel@uni-tuebingen.de}

\affiliation{Theoretical Astrophysics, Eberhard Karls University of T\"ubingen, T\"ubingen 72076, Germany}

\author{Daniela D. Doneva}
\email{daniela.doneva@uni-tuebingen.de}
\affiliation{Theoretical Astrophysics, Eberhard Karls University of T\"ubingen, T\"ubingen 72076, Germany}
\affiliation{INRNE - Bulgarian Academy of Sciences, 1784  Sofia, Bulgaria}

\author{Stoytcho S. Yazadjiev}
\email{yazad@phys.uni-sofia.bg}
\affiliation{Theoretical Astrophysics, Eberhard Karls University of T\"ubingen, T\"ubingen 72076, Germany}
\affiliation{Department of Theoretical Physics, Faculty of Physics, Sofia University, Sofia 1164, Bulgaria}
\affiliation{Institute of Mathematics and Informatics, 	Bulgarian Academy of Sciences, 	Acad. G. Bonchev St. 8, Sofia 1113, Bulgaria}

\begin{abstract}
In this work we analyze some judiciously chosen solutions of Kerr Black Holes with Scalar Hair (KBHsSH) of special interest for Gravitational Wave (GW) events originated from Extreme Mass Ratio Inspirals (EMRIs). Because of the off-center distribution of energy density, these spacetimes are warped in such a way that not all metric functions behave monotonically on the equatorial plane as in the exterior region of Kerr black holes (KBHs). This has great impact on the orbital parameters, which in turn affects the imprints on signals descendant from EMRIs in a adiabatic evolution. By investigating circular obit parameters, we unveil what qualitative features could be present in the signals that are new and distinct compared to KBHs, and we evolve some inspirals by employing the usual quadrupole formula approximation. We show that the frequencies of the emitted signals behave nonmonotonically, i.e. they can backward chirp, and for some particular cases they can become arbitrarily small, falling below LISA's sensibility range. Finally, we present two sets of waveforms produced by a noncircular EMRI in which the compact object (CO) follows a type of geodesic motion typically present in spacetimes with a static ring (SR), in which the compact object is periodically momentarily at rest.
\end{abstract}

\maketitle

\section{INTRODUCTION}

Since the first detection of gravitational waves (GW) in 2015 \cite{PhysRevLett.116.061102}, the Advanced LIGO and advanced VIRGO GW detectors have reported dozens of coalescing events, opening a new window that allows us to observe the last stages of life of binary black holes, binary neutron stars and more recently neutron-star black hole binary systems \cite{PhysRevX.11.021053,Abbott_2021}. The ultimate success of this mission fueled further confidence in the future Laser Interferometer Space Antenna (LISA), designed to detect GW signals of much lower frequencies than ground based observatories are capable of. LISA is therefore expected to open windows to many different astrophyiscal phenomena so far inaccessible to us, among which extreme mass ratio inspirals (EMRIs) \cite{Barausse:2020rsu}. These comprehend a compact object (CO) inspiraling and eventually being swallowed by a black hole, or an exotic object, a few orders of magnitude more massive than it. While the fluxes are solely due the orbiting CO, the observed signal will fundamentally be characterized by the central object and the orbital path followed by the CO. Thus, it is of utmost importance to understand what kind of imprints to expect from each massive body candidate in order to tell them apart.

There exists in the literature different approaches to investigate EMRIs. The most sophisticated one involves solving the Teukolsky equation, which gives accurate enough results to build templates for LISA. This, however, is not only notoriously expensive computationally and time-wise, but also thus far limited to Kerr spacetimes. A more robust technique is the \emph{hybrid formalism} which combines exact strong-field geodesics with weak-field radiation reaction formulas, which has been applied for general motion around Kerr and quasi-Kerr spacetimes \cite{PhysRevD.66.064005,PhysRevD.66.044002,Glampedakis_2006,PhysRevD.75.024005,PhysRevD.96.044005,PhysRevLett.126.141102}, and spherically symmetric boson stars (BSs) \cite{PhysRevD.71.044015}. In \cite{Macedo_2013,PhysRevD.88.064046}, background perturbations of a point particle inspiriling down a BS have been taken into account in the overall multipolar fluxes revealing the existence of resonant frequencies.  
The study of EMRIs around spinning central objects of only numerical know solutions is still very scarce. Recently, the imprint of surface reflectivity on waveforms was investigated by considering rotating stars of exterior Kerr vacuum, by employing the Teukolsky formalism \cite{Maggio:2021uge}.

One of the major goals of LISA will be to probe the nature of the supermassive CO in the center of the galaxies through observations of EMRIs that will be ultimately a test of our understanding of the strong field regime of gravity. Different beyond-Kerr black hole alternatives were proposed in certain extensions of Einstein's theory (see e.g. \cite{Herdeiro:2015waa}). 
As particular cases of quadratic theories of gravity, dynamical-Chern-Simons (dCS) and Einstein-scalar-Gauss-Bonnet (EsGB) theories deviate from general relativity as the respective curvature invariants are coupled to a dynamical scalar field allowing the growth of black hole hair. Several EMRI investigations have been done in order to probe these theories, both with post-Newtonian expansions and the hybrid formalism. See \cite{PhysRevD.80.064006,PhysRevD.83.104048,PhysRevD.86.044010} for EMRIs within dCS theories, and \cite{PhysRevD.86.081504} for both. In the work \cite{PhysRevD.89.044026} further applications than EMRIs, such as similar stellar mass mergers is analyzed in both theories, while \cite{PhysRevD.85.064022} reported radiative effects in general quadratic theories. More recently, the presence of a gravitational scalar charge in the inspiriling CO has been considered in \cite{PhysRevLett.125.141101,Maselli:2021men}, and its imprints in the observed signal can be strong even if the central more massive object is a Kerr black hole.

An entirely different type of hairy black hole which deserves atention is the Kerr black holes with synchronized scalar hair (KBHsSH) first reported in the novel paper of Herdeiro-Radu a few years ago \cite{PhysRevLett.112.221101,Herdeiro_2015}. These are rotating black holes with solitonic hair, whose parameter space continuously connects KBHs to  BSs, thus sharing features of both types of object. Still within the context of general relativity, the discovery of such solutions amuses for it evades the Kerr hypothesis and the no-hair theorems thanks to the matter field not sharing the same isometries of the spacetime, and the presence of superradiant instability for sufficiently fast rotating black hole. The KBHsSH resembles up to a certain extend a configuration that consist of a central black hole surrounded by a boson star. Spinning BSs, though, have a topology different from $\mathcal{S}^2$. In particular, for the most fundamental mode of rotation, the scalar field contourlines has a toroidal shape, producing an off-center energy density distribution which in turn causes the metric potentials to bear local extremes in similar fashion. This feature is also present in a large portion of KBHsSH as well, whose spacetimes greatly depart from those of Kerr BHs. Hence, it is natural to expect large deviations also in astrophysical phenomena around these objects, and to that end there has been studies on their shadows \cite{Cunha:2015yba,Cunha:2016bpi}, thin accretion disks \cite{Collodel:2021gxu}, polish doughnuts \cite{Gimeno-Soler:2021ifv}, K$\alpha$ Iron lines \cite{Ni_2016}, geodesic properties \cite{Delgado:2021jxd} and their frequencies' implications in accretion phenomena \cite{PhysRevD.95.124025}, but nothing on GW emission yet. The stability, and therefore the astrophysical relevance, of such objects throughout its whole domain of existence is still an open question. All KBHsSH contain ergoregions and are thus prone to harmful superradiant instabilities similar to those that grant their hypothetical existence. Mode stabilities derived from linear perturbations indicate that these solutions are unstable, but highly sensitive to their energy scale and with instability timescales that grow with the amount of hair \cite{PhysRevLett.120.171101,DEGOLLADO2018651}, so that for the correct parameter values the decaying time surpasses the age of the Universe. We emphasize that we are interested in solutions that share most features from rotating BSs and are therefore very hairy.

In this work, we select KBHsSH solutions whose metric functions are of particular interest, and we employ the quadrupole hybrid formalism to orbits on the equatorial plane of KBHsSH. These comprehend spacetimes endowed with, or on the verge of developing a SR as described in \cite{Collodel:2017end}, and another with a \emph{Saturnlike} ergoregion. By analyzing their orbital parameters, we anticipate what traits to expect during the inspiral which we evolve for the circular case. Furthermore, a class of non-circular orbits nonexistent around Kerr BHs is considered for producing waveforms, which we qualitatively scrutinize. 

In Section \ref{sec:theory} we do a short review on KBHsSH, circular geodesics and the quadrupole formula approximation. We then combine them in Section \ref{sec:results} where we present the results of this study. Finally, we summarize in Section \ref{sec:conclusions}. Throughout this paper we assume $G=c=1$, and restore to physical units when convenient. In our notation, $\partial_\mu\equiv\partial/\partial x^\mu$, and the overdot is a total derivative with respect to the coordinate time $\dot{x}\equiv dx/dt$.

\section{THEORY}
\label{sec:theory}

\subsection{Kerr Black Holes with Scalar Hair}
The system consists of a compound configuration of black hole with gravitating solitonic hair in the context of general relativity and is described by the following action
\begin{equation}
\label{action}
S=\int \left[\frac{R}{2}-g^{\mu\nu}\partial_\mu\Phi^*\partial_\nu\Phi-2U(\Phi)\right]\sqrt{-g}d^4x, 
\end{equation}
where $R$ is the Ricci curvature, $g$ is the metric determinant and $\Phi$ is a complex scalar field whose mass and self interaction are determined by the potential $U$. In this work we shall restrict our analysis to a massive scalar field, i.e. $U=m_b\Phi\Phi^*/2$.  This action also describes pure BSs, i.e. with no null hypersurface, as well as bald black holes when the field becomes trivial. Hairy black holes, nevertheless, are only possible for rotating spacetimes.

In adapted spherical coordinates  $\{t, r, \theta, \varphi\}$, we employ the following metric Ansatz 
\be
\label{ds1}
ds^2=-\mathcal{N}e^{2F_0}dt^2+e^{2F_1}\left(\frac{dr^2}{\mathcal{N}}+d\theta^2\right)+e^{2F_2}r^2\sin^2\theta\left(d\varphi-\omega dt\right)^2,
\ee
where $\mathcal{N}\equiv 1-r_H/r$, $r_H$ is the horizon radius and $\{F_0, F_1, F_2, \omega\}$ are the metric functions we need to solve for, all dependent on both  $r$ and $\theta$, that completely describe the spacetime geometry. This spacetime is axisymmetric and stationary, and possesses therefore two Killing vector associated with these isometries, $\xi^\mu=\vec{\partial}_t$ and $\chi^\mu=\vec{\partial}_\varphi$. Note that the usual Kerr solution in Boyer-Lindquist coordinates has a different parametrization since its radial coordinate $\bar{r}$ is not the same, but related via
\be
\label{eq:kerrtor}
\bar{r}=r+\frac{a^2}{\bar{r}_H}, \qquad \bar{r}_H=M+\sqrt{M^2-a^2}.
\ee
Upon comparing hairy and bald solutions we shall use $r$ since it is possible to write KBHs in this form, but not KBHsSH. The reason is that we can have overspinning KBHsSH with $a>M$, and furthermore there is no uniqueness for a fixed pair of \{M,a\} \cite{PhysRevLett.112.221101}.

The underlying scalar field theory is endowed with a Noether symmetry, since the global $U(1)$ transformation $\Phi\rightarrow e^{i\alpha}\Phi$ leaves the system invariant. The associated conserved current and charge are given by,
\begin{equation}
\label{ncharge}
j^\mu=-i\left(\Phi^*\partial^\mu\Phi-\Phi\partial^\mu\Phi^*\right), \qquad Q=\int_{\Sigma\backslash\mathcal{H}}j^\mu n_\mu dV,
\end{equation}
where $\Sigma$ is the three-space hypersurface, $\mathcal{H}$ is the horizon volume, $dV$ is the volume element, $n^\mu$ is the unit timelike vector orthogonal to $\Sigma$, such that $n_\mu=(-   e^{F_0}\sqrt{\mathcal{N}},0,0,0)$.

Bound state configurations of the scalar field theory at hand have their stability properties intrinsically connected to the conserved charge above. It is straightforward to see that the static field possesses zero net charge, and hence the scalar field must be time dependent. Similarly, rotating solitons need also to be dependent on the axial coordinate in order to have nontrivial momentum in this direction. This is the case for solitons in flat spacetime, as well as for gravitating solitons potentially combined with a black hole \cite{PhysRevD.66.085003,Yoshida:1997qf,Mielke2016,PhysRevD.72.064002,PhysRevD.101.044021,PhysRevLett.112.221101,PhysRevD.102.084032}. Thus, the scalar field depends on all four spacetime coordinates, but since the metric is stationary and axisymmetric, this dependence must be explicitly harmonically in $t$ and $\varphi$ for consistency, as 
\begin{equation}
\Phi=\phi(r,\theta)e^{i(\omega_st+m\varphi)},
\end{equation}
where $m$ is the integer winding number and $\omega_s$ is its natural frequency. 

Hairy solutions do not occur for any values of the input parameters $\{r_H,\omega_s,m\}$, for it is consequence of a superradiance phenomenon. In the linearized regime, i.e. solving the Klein-Gordon equation on a Kerr background, a transition to superradiance instability  happens when the horizon angular velocity matches the angular velocity of the scalar field, $\omega_H=\omega_s/m$. In the full nonlinear regime, this relation arises as a regularity condition at the horizon. Hence, the hair is \emph{synchronized} with the hole. 

BSs have the particular trait that their angular momentum is quantized, and KBHsSH feature the same property for the angular momentum stored in their hair, $J_\Phi=mQ$. Since the total angular momentum is the sum of the contributions from the hole and the hair $J=J_H+J_\Phi$, a natural way of assessing how hairy a solution is is given by the dimensionless normalized charge $q\equiv J_\Phi/J$, so that $q=0$ for KBHs and $q=1$ for BSs.

All solutions used in this work are in their fundamental mode of rotation, so $m=1$. The boson mass, $m_b$, is absorbed by the radial coordinate and the field's frequency such that the following quantities rescale as 
\begin{equation}
\label{mus}
r\rightarrow rm_b, \qquad \omega_s\rightarrow \omega_s/m_b, \qquad M\rightarrow Mm_b.
\end{equation}

\begin{figure}[htp]
\centering
\includegraphics[width=.4\textwidth]{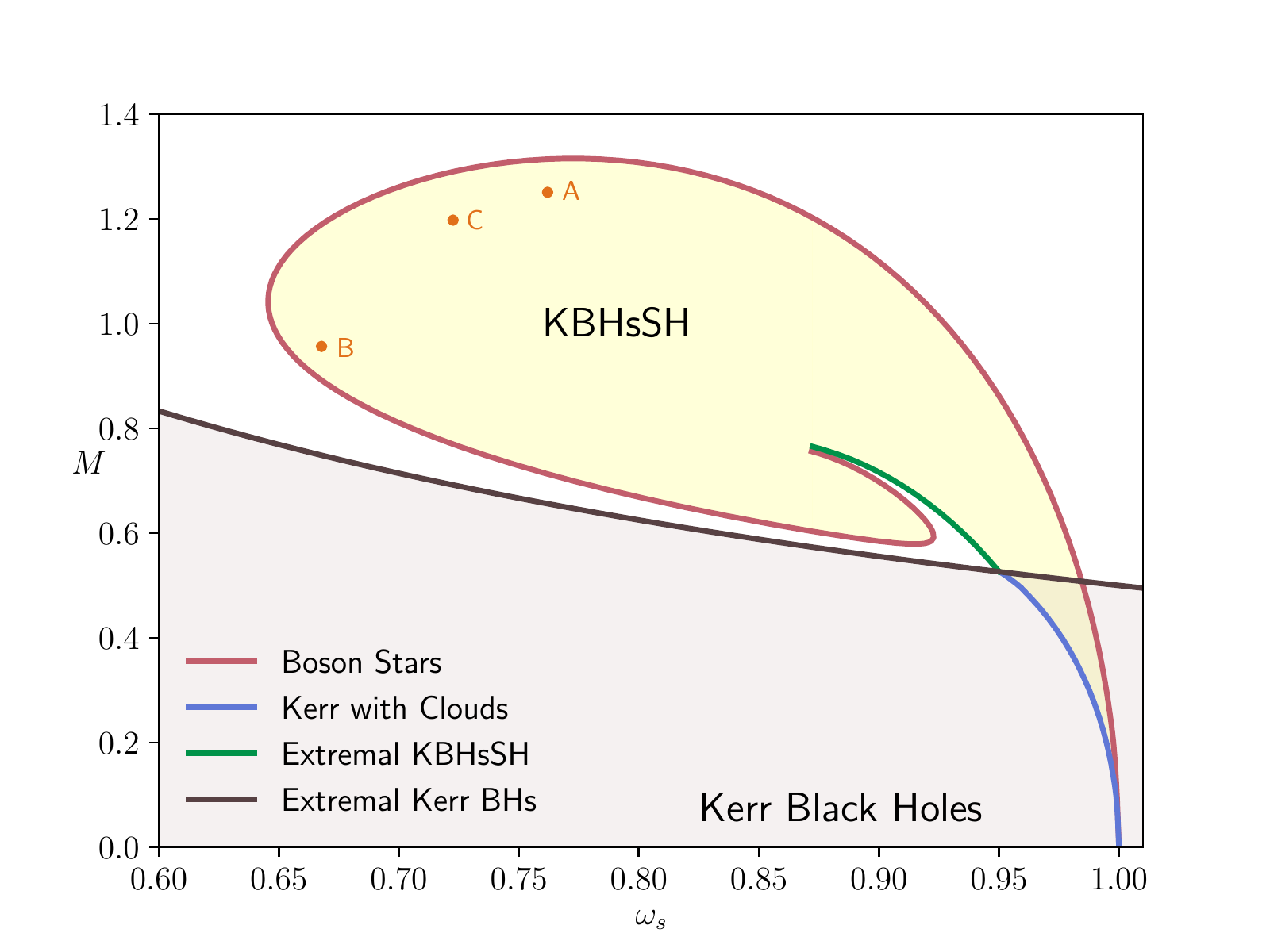}
\caption{Domain of existence of KBHsSH, which is bounded by Kerr BHs with scalar cloud (blue), BSs (red) and extremal KBHsSH (green). Orange dots with associated letters indicate the solutions which are used in Section \ref{sec:results}.}
\label{fig:domain}
\end{figure}

The domain of existence of KBHsSH is shown in Fig. \ref{fig:domain} in an $\omega_s$ vs. $M$ diagram, together with the particular solutions we work with in the next section. The domain is bounded by three curves of solutions, each with its specific properties. In blue we have the probe limit, i.e. Kerr BHs with non-backreacting scalar clouds ($q=0$), in red are the BS solutions ($q=1, r_H=0$), and in green extremal KBHsSH ($r_H=0$).

\subsection{Orbits on the Equatorial Plane}

The orbits here considered are restricted to motion on the equatorial plane, $d\theta=0$. Stationarity and axisymmetry grant two first integral of motion for an orbiting particle or CO around the black hole, namely the energy $E=-u_t$ and $L=u_t$, respectively, where we normalized with respect to the mass of the orbiting object. The equation of motion in the radial direction is taken from the norm of the normalized four-velocity $u^{\mu}u_{\mu}=-1$,
\begin{equation}
\label{eq:rd}
V_{eff}\equiv g_{rr}\left(\frac{dr}{d\tau}\right)^2=\frac{g_{\varphi\varphi}}{g_{t\varphi}^2-g_{tt}g_{\varphi\varphi}}\left(E-V_+\right)\left(E-V_-\right),
\end{equation}
where we defined and effective potential, and
\begin{equation}
\label{vpm}
V_\pm\equiv L\frac{g_{t\varphi}}{g_{\phi\phi}}\pm\frac{\sqrt{L^2\left(g_{t\varphi}^2-g_{tt}g_{\varphi\varphi}\right)}}{g_{\phi\phi}}.
\end{equation}

The angular velocity is defined as $\Omega\equiv u^\varphi/u^t=\dot{\varphi}$. For circular orbits, $V_{eff}=dr/d\tau=d^2r/d\tau^2=0$, Then the orbital parameters, $\{E,L,\Omega\}$ are given in general form by
\begin{equation}
\label{eqen}
E=-\frac{g_{tt}+g_{t\varphi}\Omega}{\sqrt{-g_{tt}-2g_{t\varphi}\Omega-g_{\varphi\varphi}\Omega^2}},
\end{equation}
\begin{equation}
\label{eqan}
L=\frac{g_{t\varphi}+g_{\varphi\varphi}\Omega}{\sqrt{-g_{tt}-2g_{t\varphi}\Omega-g_{\varphi\varphi}\Omega^2}},
\end{equation}
\begin{equation}
\label{eq:om}
\Omega_\pm=\frac{-\partial_rg_{t\varphi}\pm\sqrt{(\partial_rg_{t\varphi})^2-\partial_rg_{tt}\partial_rg_{\varphi\varphi}}}{\partial_rg_{\varphi\varphi}}.
\end{equation}

The innermost circular orbit (ISCO) dwells at smallest radius for which the orbit is marginally stable, meaning $\partial^2_rV_{eff}=0$. Thus, it is found for the smallest radius $r=r_{ISCO}$ that satisfies the following equality,
\begin{equation}
\label{vrr}
E^2\partial_r^2g_{\varphi\varphi}+2EL\partial^2_rg_{t\varphi}+L^2\partial_r^2g_{tt}-\partial^2_r\left(g_{t\varphi}^2-g_{tt}g_{\varphi\varphi}\right)=0.
\end{equation}

We remark that unlike circular geodesics on the equatorial plane of Kerr BHs, for several KBHsSH there exists more than one region of stable and unstable orbits \cite{Collodel:2021gxu,Delgado:2021jxd}. Hence, we define the outermost unstable circular orbit (OUCO) as the one with the largest radius to satisfy eq. (\ref{vrr}).

Some of these spacetimes are endowed with a SR, i.e. a ring of points where an orbiting particle is permanently at rest with respect to a zero angular momentum observer at infinity, or a fiducial observer. It is simple to see, from eq. (\ref{eq:om}) that the necessary condition to form the appropriate stage for it, is the presence of a local extreme in the $g_{tt}$ component of the metric outside of an ergoregion, so that $\Omega_-=0$. Most solutions for which this happens contain in reality two distinct SRs due to the presence of a local minimum and a local maximum. In the region delimited by the SRs, $\Omega_->0$ and therefore both orbits therein are prograde but of different magnitudes for the orbital velocity, and furthermore both might be stable. Thus, it is senseless to directly associate $\Omega_+$ and $\Omega_-$ with co-rotating and counter-rotating orbits respectively.

Noncircular geodesics are found by solving eq. (\ref{eq:rd}) with the appropriate initial conditions. In particular, we are interested in a class of orbits absent around KBHs, in which the particle or CO is periodically instantaneously at rest. This is always possible if the spacetime features a SR. A particle initially at rest at a radius beyond the one that defines the SR $r_{sr}$ will counter-rotate with respect to the black hole in an orbit called \emph{pointy-petal} due to the motion it describes. If, however, it starts initially at rest between the hole and $r_{SR}$, it will co-rotate in a \emph{semi}-orbit \cite{Collodel:2017end,PhysRevD.90.024068,Grould_2017}. 

\subsection{EMRIs} 
\label{ssec:emri}

The most prominent contribution for GW produced in an EMRI comes from the $\ell=2$ mode. Neglecting higher modes, the mass moments reduce to the quadrupole moment tensor \cite{Thorne:1980ru} 
\be
\label{eq:Istf}
\mathcal{I}_{ij}=\left[\int \rho x_ix_jd^3x\right]^{\textrm{STF}},
\ee
where STF means we cast the tensor in a symmetric trace-free form and $\rho=\mu\delta^3(\mathbf{x}-\mathbf{x}_\mathrm{CO})$ and $\mu$ is the CO's mass normalized by the central object's mass. This tensor is evaluated in flat-space such that
\be
x_1=x=r\cos\varphi, \qquad x_2=y=r\sin\varphi, \qquad x_3=z=0,
\ee
and we recall that the motion is restricted to the equatorial plane. The fluxes are then given by,
\be
\label{eq:fen1}
\mu\dot{E}=\frac{1}{5}\dddot{\mathcal{I}}_{ij}\dddot{\mathcal{I}}_{ij},
\ee
\be
\label{eq:fan1}
\mu\dot{L}_k=\frac{2}{5}\epsilon_{kij}\ddot{\mathcal{I}}_{ij}\dddot{\mathcal{I}}_{ij}.
\ee

For circular orbits, equations (\ref{eq:fen1}), (\ref{eq:fan1}) together with (\ref{eq:Istf}) obviously yield the same results as taking the time derivative of eq. (\ref{eqen}) and (\ref{eqan}),
\be
\label{eq:fen2}
\mu\dot{E}=\frac{32}{5}\mu^2r^4\Omega^6, \qquad \mu\dot{L}_k=\frac{1}{\Omega}\frac{d(\mu E)}{dt}.
\ee
Note that these are the fluxes measured at infinity and therefore the loss of energy and angular momentum by the CO should be minus these quantities.

The gravitational wave field is then given simply by
\be
\label{eq:hijtt}
h_{ij}=\frac{2}{D}\ddot{\mathcal{I}}_{ij},
\ee
where $D$ is the distance between the compact object and the observer. However, the observable part is given by the transverse traceless (TT) part of this tensor,
\be
h_{ij}^{\textrm{TT}}=P_{ik}h_{kl}P_{lj}-\frac{1}{2}P_{ij}P_{kl}h_{kl},
\ee
for which $P_{ij}=\delta_{ij}-n_in_j$ is the projection operator and $n_i$ is the unit vector pointing from the observer to the source. The polarization tensors $H_{ij}^+$ and $H_{ij}^\times$ read
\be
\label{eq:ptensors}
H_{ij}^+=p_ip_j-q_iq_j, \qquad H_{ij}^\times=p_iq_j+q_ip_j,
\ee
in terms of the unit vectors
\be
\label{eq:uvectors}
p_i\equiv\frac{\epsilon_{ijk}n_jL_k}{|\epsilon_{ijk}n_jL_k|}, \qquad q_i\equiv\epsilon_{ijk}n_jp_k.
\ee

The GW field can thus be expanded into the polarization modes, such as
\be
\label{eq:hijtt+x}
h_{ij}^{\textrm{TT}}=A_+(t)H_{ij}^+ +A_\times(t)H_{ij}^\times,
\ee
and the wave's amplitudes can be expressed by
\be
\label{eq:amplitudes_general}
A_+(t)=\frac{1}{2}H_{ij}^+h_{ij}^{\textrm{TT}}, \qquad A_\times(t)=\frac{1}{2}H_{ij}^\times h_{ij}^{\textrm{TT}},
\ee
which for orbits on the equatorial plane become
\be
\label{eq:a+}
A_+=\frac{2\mu}{D}\left\{ \left[ \Omega^2r^2\cos\left(2\varphi\right)+\frac{1}{2}\left(4\Omega \dot{r}+\dot{\Omega}r\right)r\sin\left(2\varphi\right) \right]\left(1+\cos^2\Theta\right)+\left(\ddot{r}r+\dot{r}^2\right)\left[\sin^2\varphi-\cos^2\varphi\cos^2\Theta\right]\right\},
\ee
\be
\label{eq:ax}
A_\times=\frac{2\mu\cos\Theta}{D}\left[ \left(\ddot{r}r+\dot{r}^2-2\Omega^2r^2\right)\sin\left(2\varphi\right)+\left(4\Omega \dot{r}+\dot{\Omega}r\right)r\cos\left(2\varphi\right)\right],
\ee
where $\Theta$ is the inclination of the equatorial plane with respect to the observer and $\phi_0$ is simply a phase constant. 

Such approach is quite simplified in the sense that it ignores the fluxes across the horizon and also higher modes. Nevertheless, it catches the most prominent features of an EMRI which can then be used as guidelines for choosing specific solutions to be used with more sophisticated methods. These methods, such as solving the corresponding Teukolsky equation, are not only extremely expensive computationally but currently pose notorious difficulties when applied to rotating spacetimes which are only prescribed numerically. 
So far we are yet to see a breakthrough in the field that will allow us to build GW templates for EMRIs in more general spacetimes. However, for face-on configuration ($\Theta=0$), simulations in KBHs show a good agreement between the waveforms produced with the quadrupole formula approximation and the Teukolsky approach, mainly presenting differences in the phase and amplitude factor. Hence, in what follows we shall consider strictly this configuration. The evolution of the norm of the amplitudes over a large timespan all the way to the plunging, on the other hand, shows even different qualitative behaviors for KBHs near extremality when performed with these two methods.

\section{Case Studies}
\label{sec:results}
In what follows, we present three distinct solutions of special interest to investigate their circular orbit structure and evolve the fluxes when convenient. In the strong field, these solutions -- which are highlighted in Fig. \ref{fig:domain} -- depart considerably from Kerr and possess very different properties which are directly linked to their $g_{tt}$ component as explained below. Since one solution can be continuously deformed into another, it is more instructive to focus on these cases to highlight what peculiar traits this region of the parameter space might hold, rather than blindly exhausting the full set of solutions. As mentioned in the introduction, we choose one solution which is very close to developing a SR ($\Omega_-$ gets very close to zero in a certain region), one with the same horizon radius that contains a pair of SRs and another one with two different ergoregions on the equatorial plane. In order to compare each hairy black hole with a Kerr black hole, we shall fix the mass and horizon radius. Contrary to the usual approach of fixing the mass and angular momentum, rendering solutions of equal spin parameter, this is better suited to compare distances from the hole. The spin parameter for the Kerr counterpart, $a_K$ is then given by
\be
\label{eq:ak}
a_{\textrm{Kerr}}=\frac{\sqrt{4M^2-r_H^2}}{2}.
\ee

In Table \ref{table:cases} we present the parameters that define each of the solutions used, together with the corresponding Kerr spin parameter.

\begin{table}[h!]
\centering
\begin{tabular}{ |*{8}{c|}}
 \cline{2-8}
  \multicolumn{1}{c|}{}      & $r_H$ & $M$ & $J$ & $\omega_s$ & $q$ & $a_{\textrm{Kerr}}$ & $a_{\textrm{Kerr}}/M$ \\
 \hline
 \hline
 \multicolumn{1}{|c||}{Case A} & $0.07$ & $1.25$ & $1.25$ & $0.76$ & $0.9992$ & $1.25$ & $0.99961$ \\ 
 \multicolumn{1}{|c||}{Case B} & $0.02$ & $0.96$ & $0.83$ & $0.67$ & $0.9996$ & $0.96$  & $0.99995$ \\ 
 \multicolumn{1}{|c||}{Case C} & $0.07$ & $1.20$ & $1.17$ & $0.72$ & $0.9984$ & $1.20$  & $0.99957$ \\ 
 \hline
\end{tabular}
\caption{Parameters defining each solution case.}
\label{table:cases}
\end{table}

\begin{figure}[htp]
\centering
\includegraphics[width=.4\textwidth]{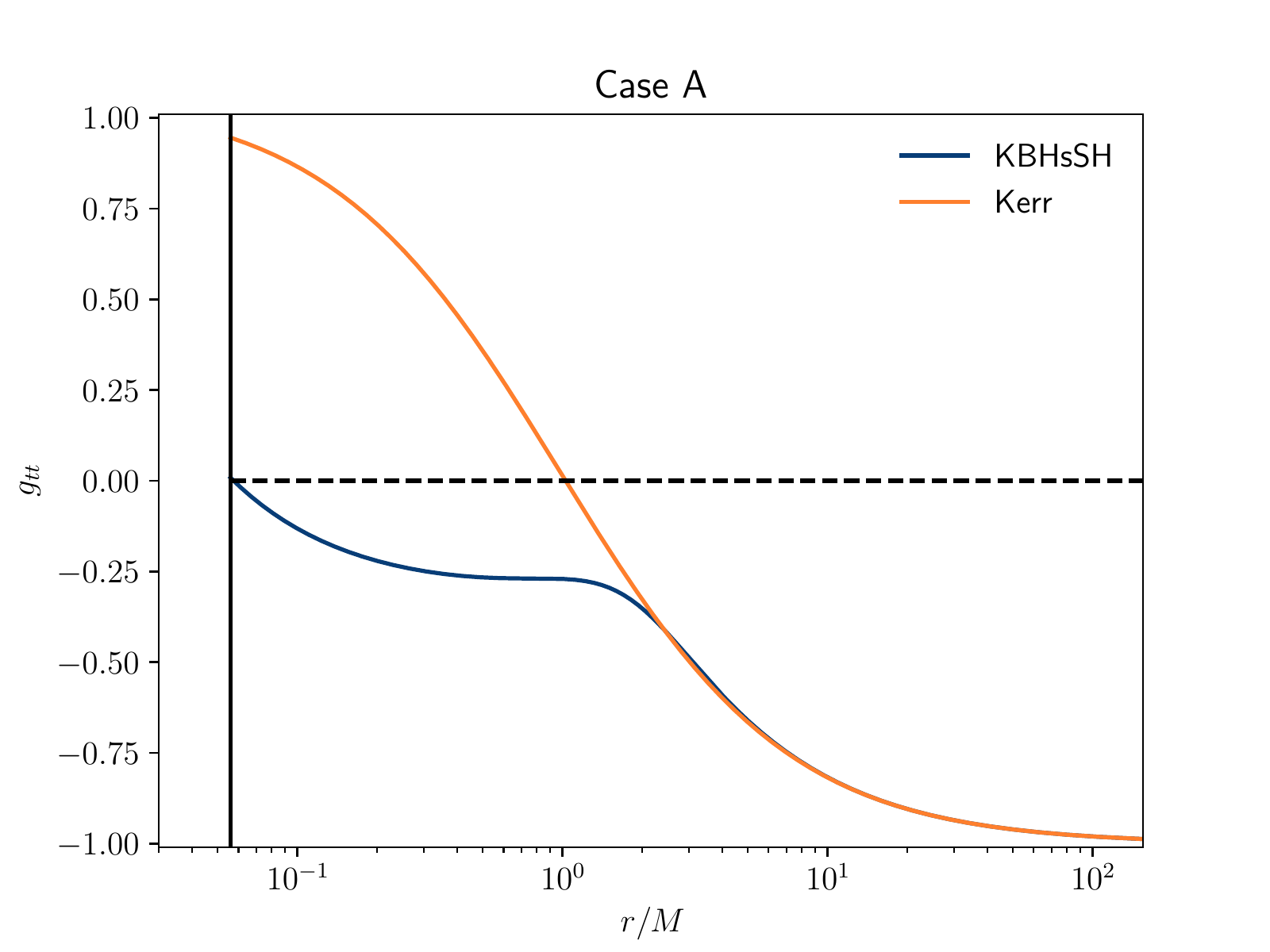}
\includegraphics[width=.4\textwidth]{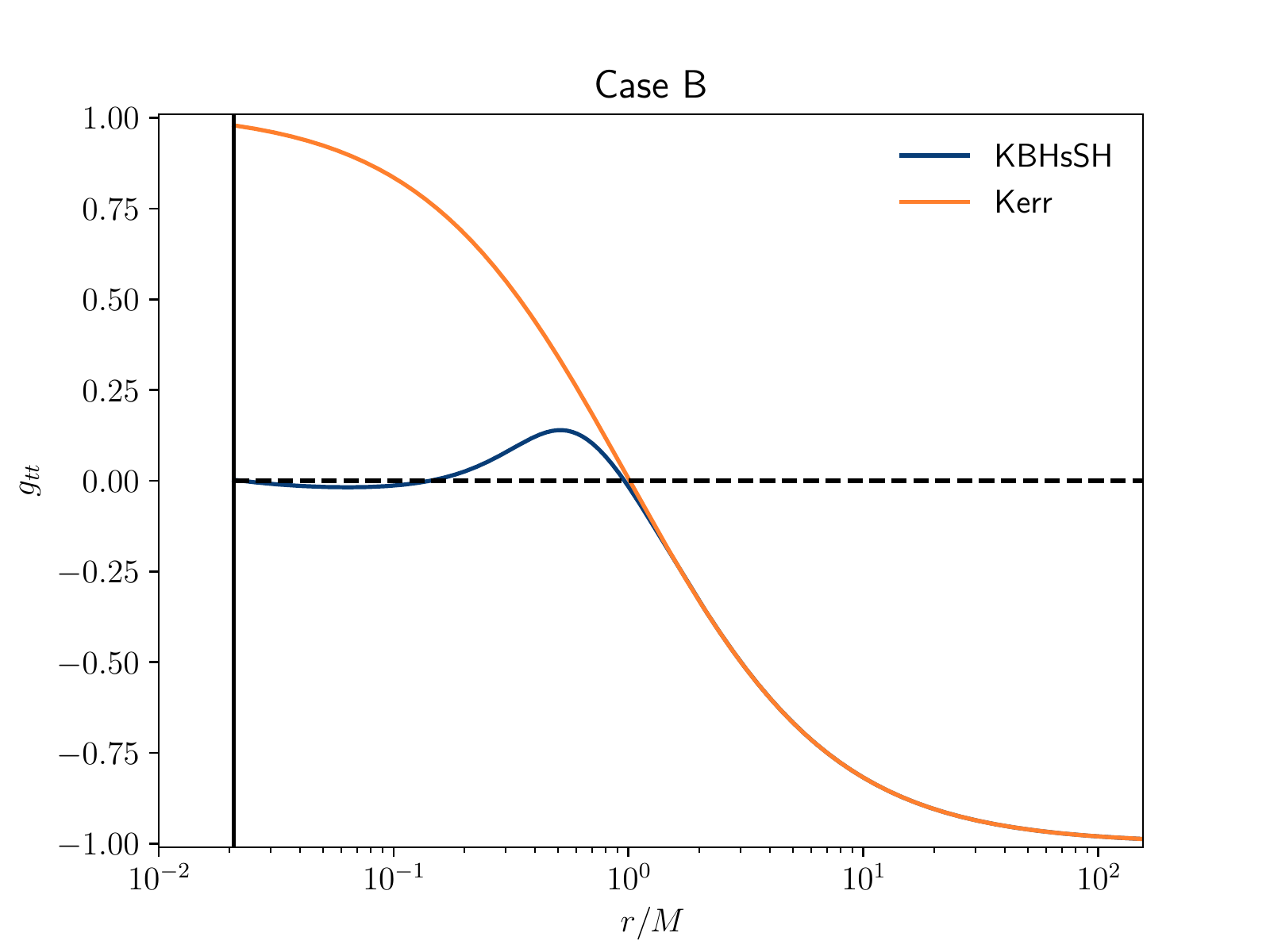} \\
\includegraphics[width=.4\textwidth]{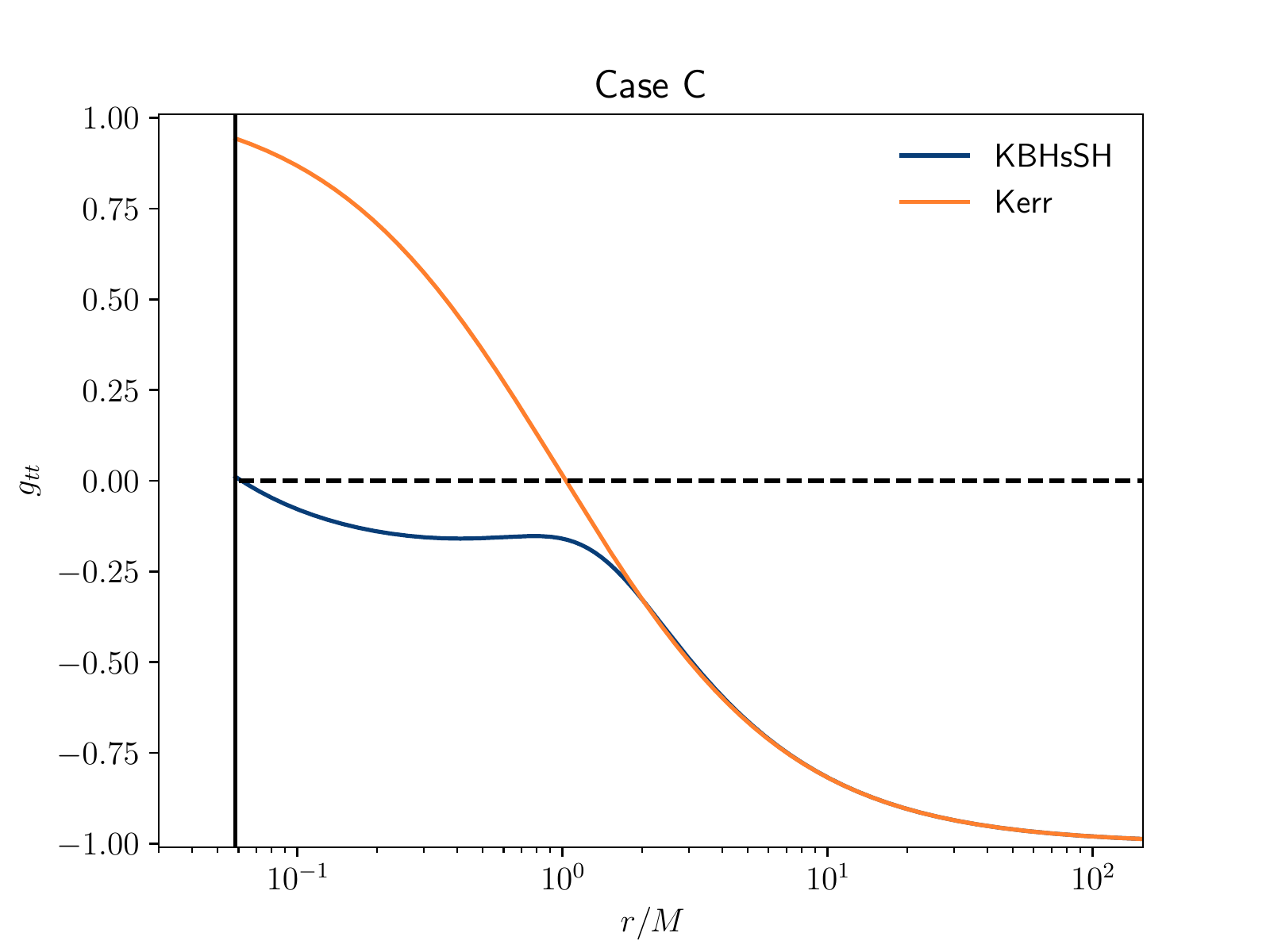}
\caption{The $g_{tt}$ component of the metric against the radial coordinate for the three particular hairy solutions and their bald counterparts of same $(M,r_H)$ parameter pair. The vertical black line indicates the event horizon position. \emph{Top Left:} Case A. \emph{Top Right:} Case B. \emph{Bottom:} Case C. 
}
\label{fig:gtt}
\end{figure}

The $g_{tt}$ component of the metric of each solution is drawn in Fig. \ref{fig:gtt} together with their Kerr counterparts. The vertical solid black line indicates the location of the horizon. We note that for large distances, there is an effective exterior region for the hairy solutions where the metric approaches that of Kerr very quickly. Inside the solitonic hair, however, the solution departs drastically from that of a bald black hole. Case A almost forms a saddle point in the $g_{tt}$ profile, which in turn will result in retrograde orbits with very small absolute orbital velocity around these points. Case B, on the other hand, features a local maximum and a local minimum for this metric function, where the retrograde orbital velocity would become zero, forming a SR. However, the maximum is located within an ergoregion and the local minimum hosts an unstable orbit. Case C is depicted in the lower panel of the figure and it features both a local maximum and minimum out of an ergoregion, forming two SRs.

\subsection{Case A}

\begin{figure}[htp]
\centering
\includegraphics[width=.4\textwidth]{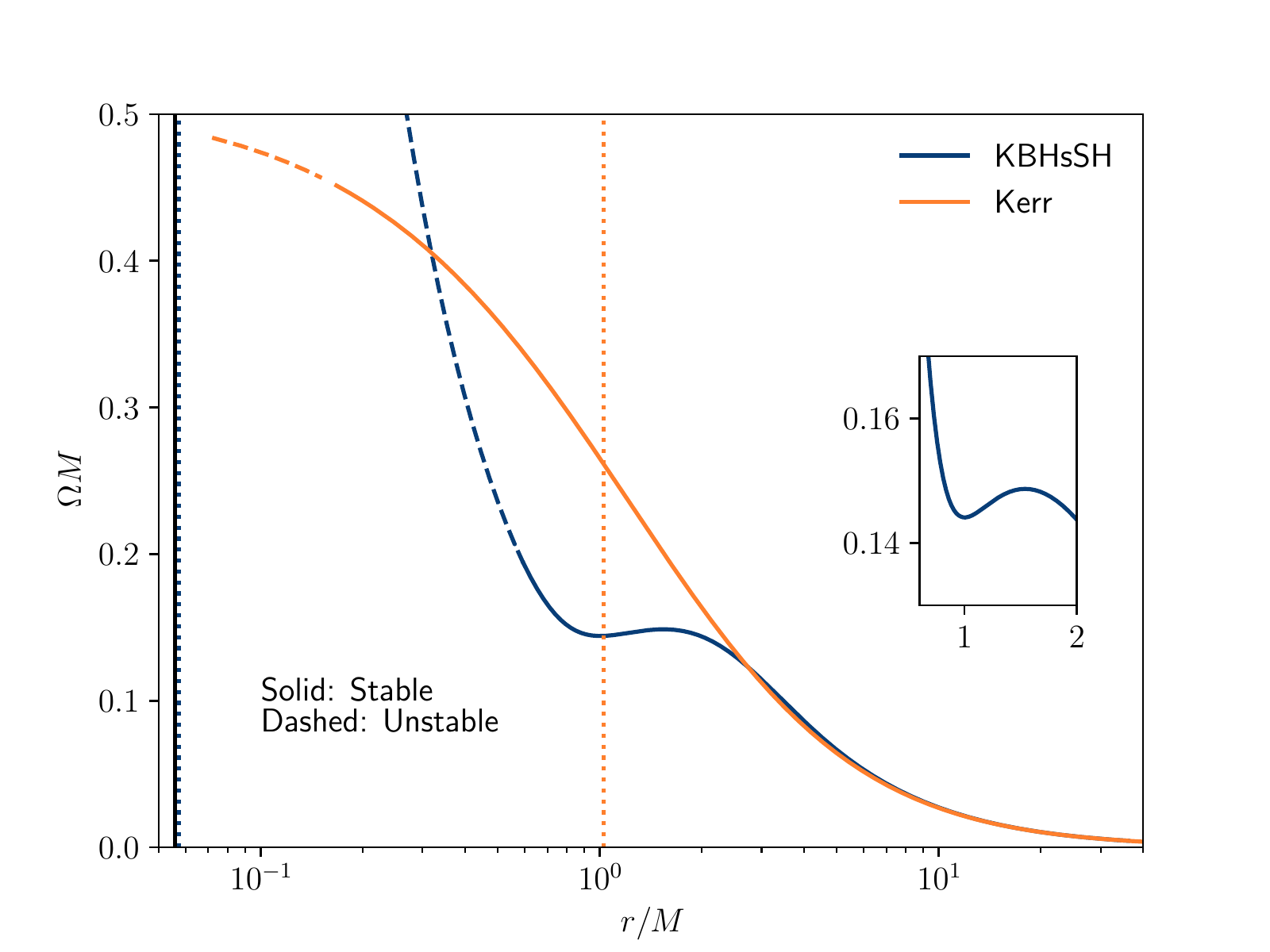}
\includegraphics[width=.4\textwidth]{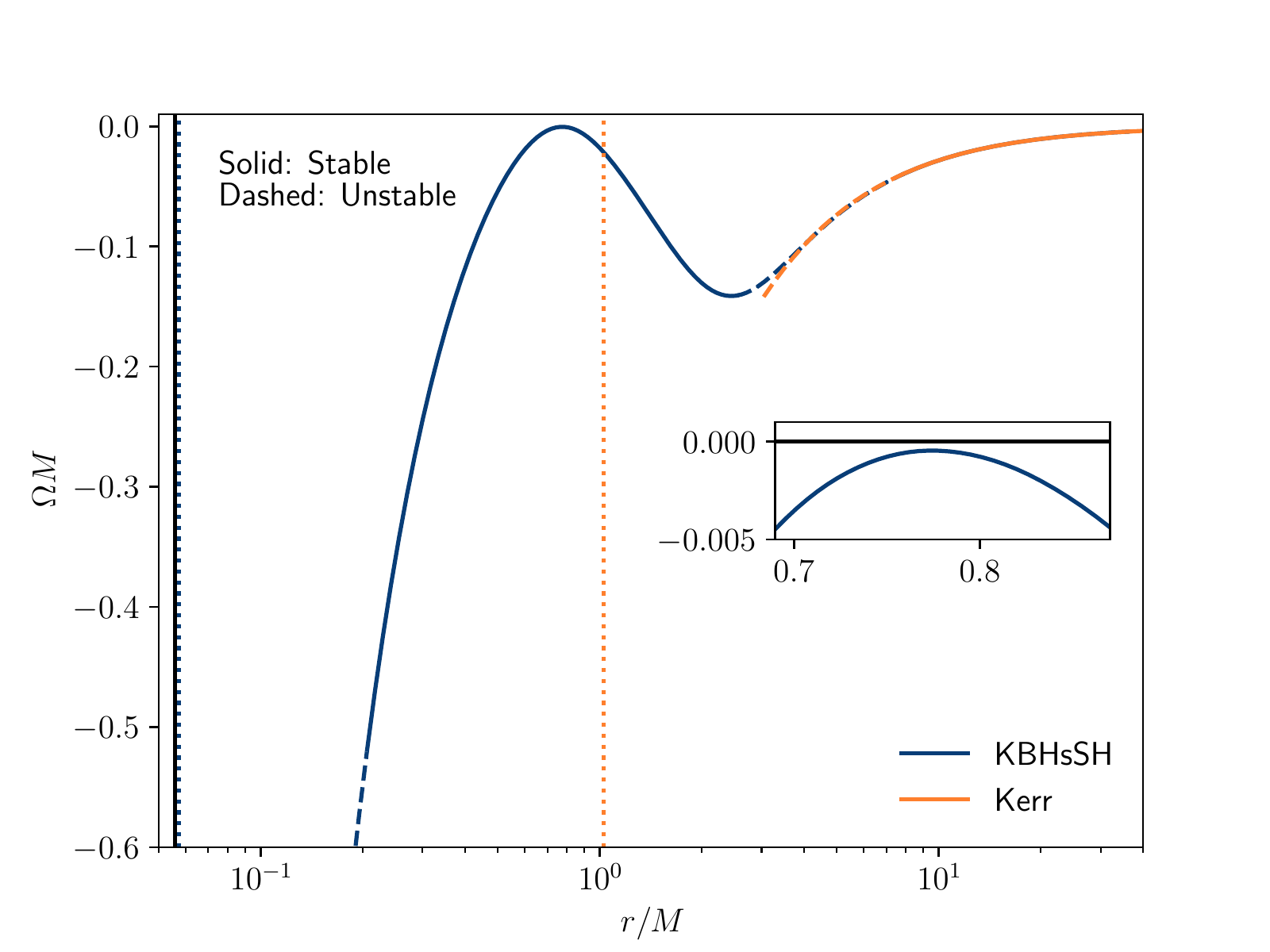}
\caption{Circular orbital velocities for Case A. \emph{Left:} $\Omega_+$. \emph{Right:} $\Omega_-$. The black vertical line represents the horizon while the dotted vertical lines correspond to the ergosurfaces.}
\label{fig:rxOas}
\end{figure}

This solution dwells very close in the parameter space to solutions containing a SR. As we have shown above, its $g_{tt}$ component almost forms a saddle point. Even though its profile still varies monotonically, that is not true for the circular orbital velocities. In Fig. \ref{fig:rxOas} we show the behavior of $\Omega_+$ and $\Omega_-$ (which here indeed correspond to prograde and retrograde orbits) with respect to the radial coordinate. In both cases, as one approaches the hole, there is a region of counter chirping, i.e. where the absolute value of the orbital velocity decreases, as opposed to what happens around bald Kerr black holes. In the prograde case, all of the circular orbits are stable after the ISCO, but in the retrograde case there are two regions of instability. As one decreases the radius, the orbits of hairy and bald black holes become unstable at very nearby points, $r_{\textrm{OUCO}}=9.86$ and $r_{\textrm{ISCO}}=10.0$, respectively. Thereon, timelike circular orbits soon cease to exist in the Kerr case, which has an ergosurface of much larger radius than the hairy black hole. Near this point, stability is regained in the KBHsSH case and the counter chirp begins, with the absolute value of the orbital velocity reaching a maximum very close to zero. We shall point out that although our solution set is discrete, they are continuously connected, and there should be therefore solutions that come arbitrarily close to forming a SR. As we keep decreasing $r$ after the maximum of $\Omega_-$, it decreases monotonically until the ISCO.

\begin{figure}[htp]
\centering
\includegraphics[width=.4\textwidth]{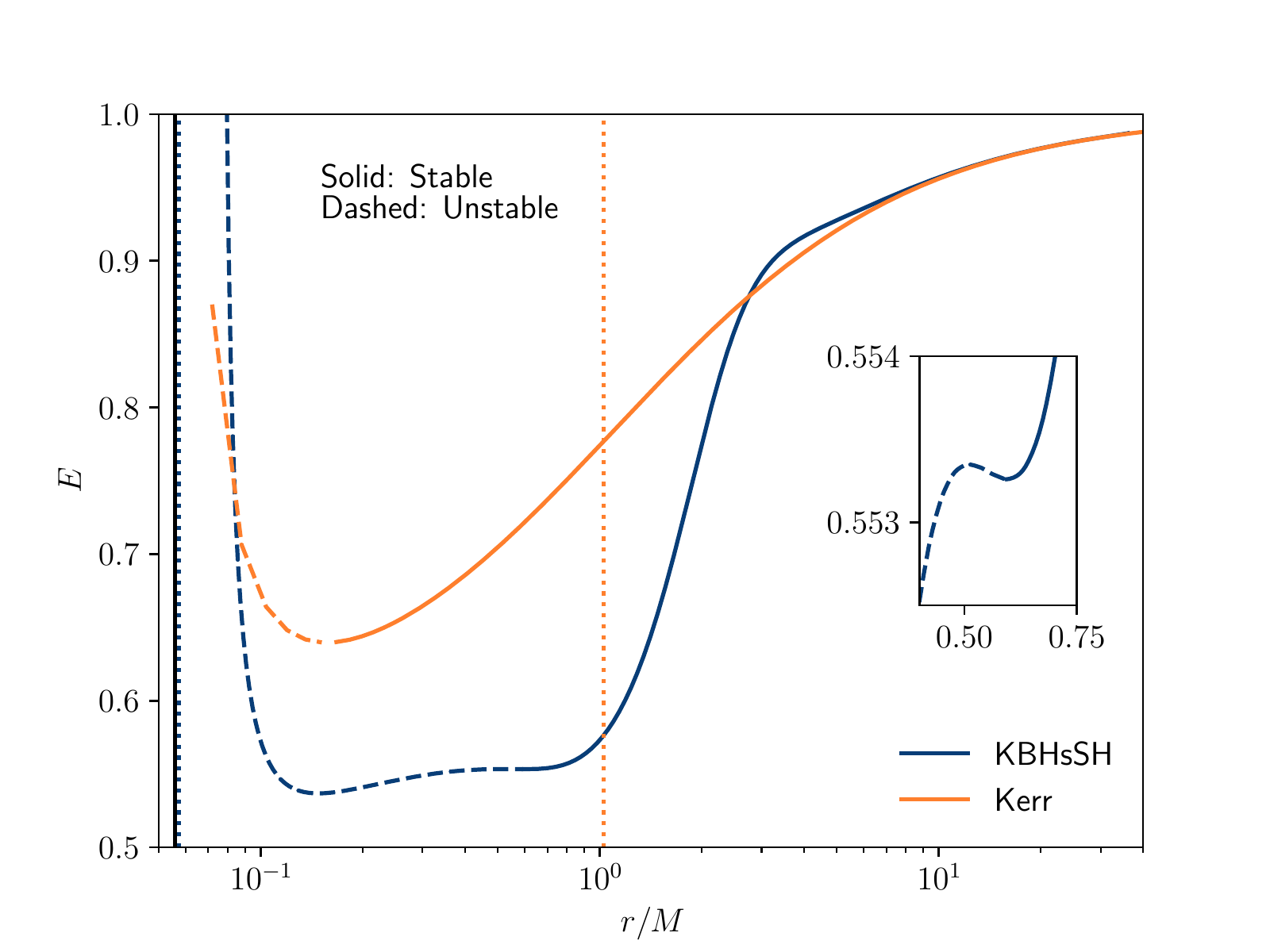}
\includegraphics[width=.4\textwidth]{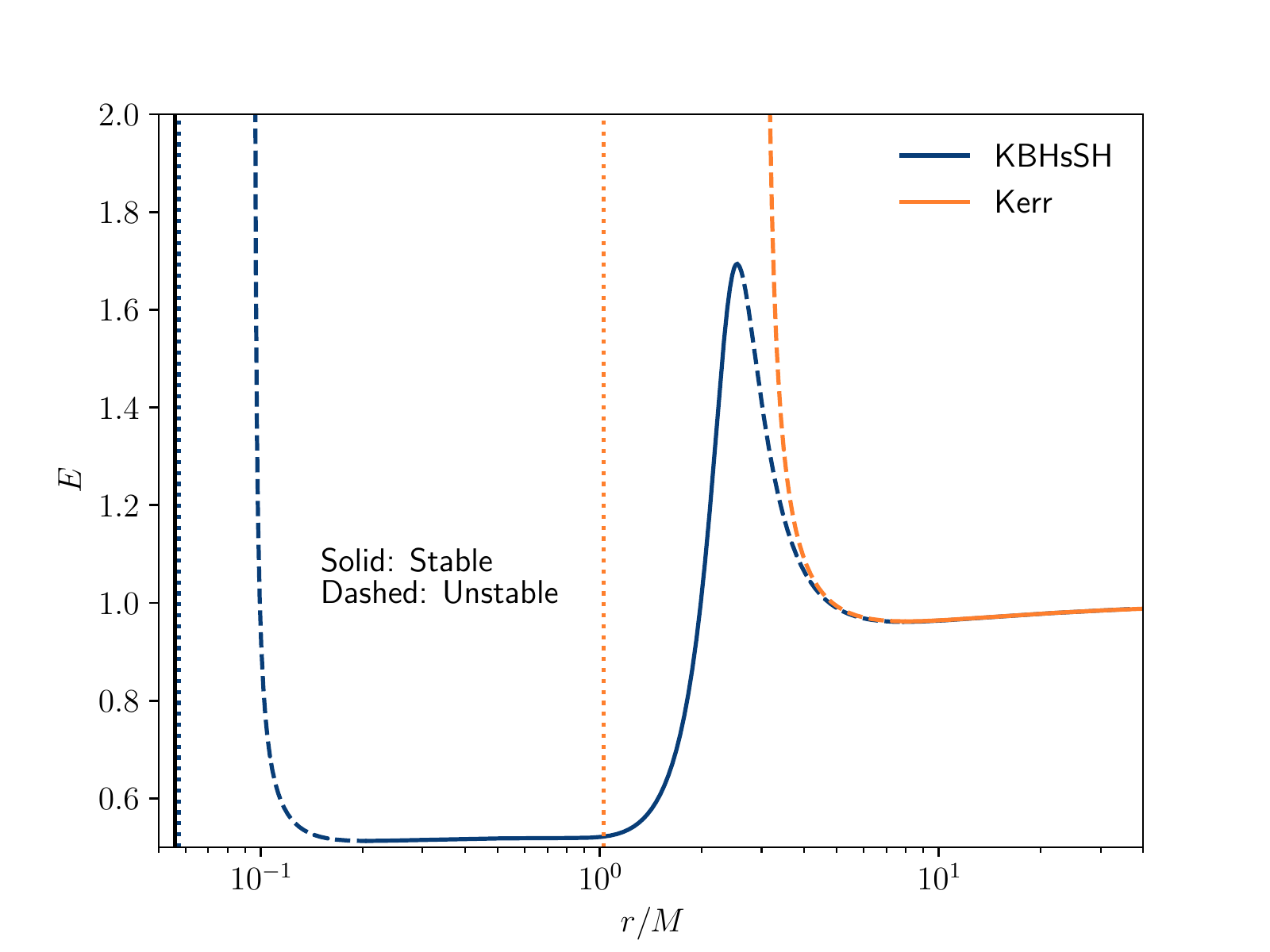}
\caption{Circular orbital energies for Case A. \emph{Left:} $\Omega_+$. \emph{Right:} $\Omega_-$. The black vertical line represents the horizon while the dotted vertical lines correspond to the ergosurfaces.}
\label{fig:rxEas}
\end{figure}

The energy profiles for the orbits discussed above are shown in Fig. \ref{fig:rxEas}. We note that for both hairy and bald solutions, the instabilities arise when the energy starts growing with decreasing $r$, although this is not always the case for KBHsSH. This illustrates well  the plunging phenomenon after the ISCO, for if the compact object were to keep approaching the hole within circular orbits it would need to gain energy.

\subsection{Case B}

\begin{figure}[htp]
\centering
\includegraphics[width=.4\textwidth]{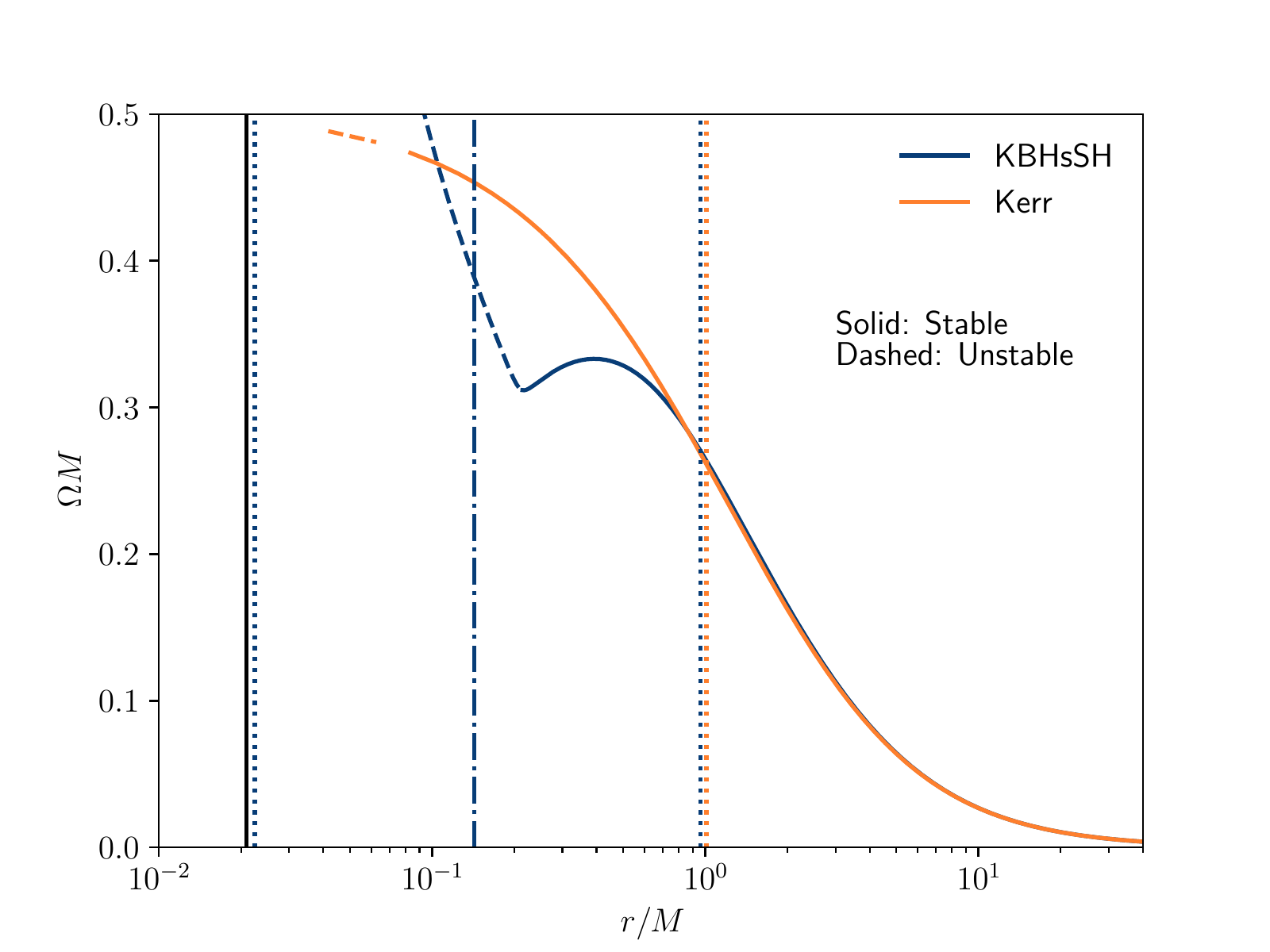}
\includegraphics[width=.4\textwidth]{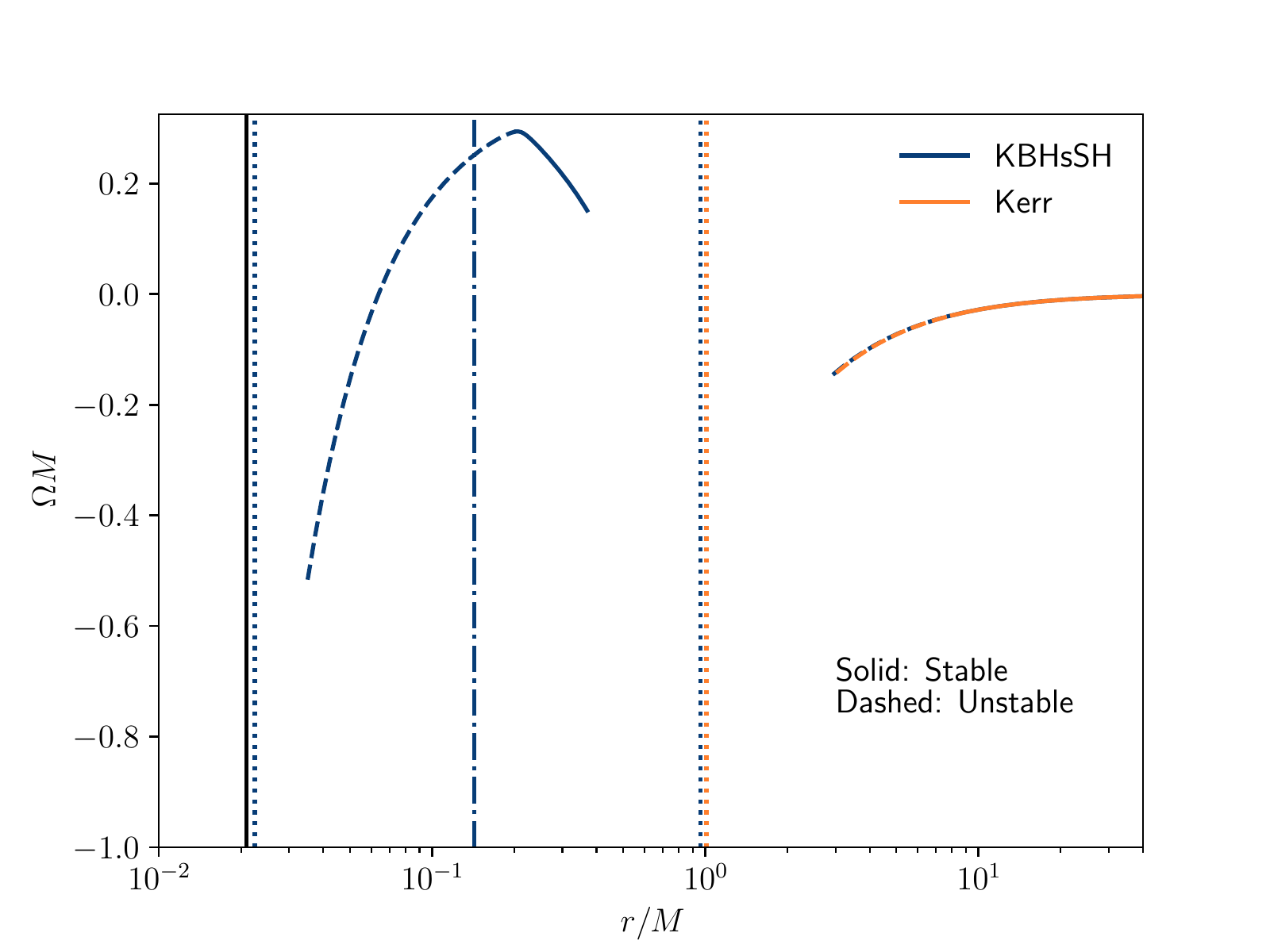}
\caption{Circular orbital velocities for Case B. \emph{Left:} $\Omega_+$. \emph{Right:} $\Omega_-$. The black vertical line represents the horizon while the dotted vertical lines correspond to the ergosurfaces. The ergoregion of the hairy solution is of the \emph{Saturn} type, meaning there are two disconnected regions. The dot-dashed vertical line corresponds to the innersurface of the outer ergoregion, and the world line of fiducial observers between the innermost dotted blue line and the dot-dashed one are thus timelike. Within the exterior ergoregion there are both stable and unstable orbits for $\Omega_-$, which is positive in this region, and there are therefore two types of prograde circular orbits within an interval of $r$.}
\label{fig:rxOrb}
\end{figure}

In Case B, the main differences arise for orbits characterized by the $\Omega_-$ eigenvalue. In Fig. \ref{fig:rxOrb} we depict the orbital velocities as before, for both $\Omega_+$ and $\Omega_-$. Nonetheless, while $\Omega_+$ always corresponds to prograde orbits, $\Omega_-$ can be both retrograde or prograde, depending on the radius of the orbit. Overall, the same counter chirping behavior can be seen in both types of orbits as before, and the main difference in the $\Omega_+$ case lies on the fact that the orbital velocity would start to grow again with decreasing radius right at the ISCO. This hairy solution contains two disconnected regions where $g_{tt}>0$, forming a \emph{Saturnlike} ergoregion. In the figure, the dot-dashed blue vertical line indicates the inner surface of the outer ergoregion, and $g_{tt}<0$ between the first dotted line and the dot-dashed one. As before, there are two regions of unstable orbits for the hairy solution. 
In between, we find both a region with no timelike circular orbits and one of stable ones entirely inside an ergoregion where $\Omega_->0$. 
The outer ergosurface almost superposes that of the Kerr solution. For larger radii we enter the effective exterior region where the geometry negligibly deviates from Kerr and the orbital parameters differ minimally. For instance, the ISCO of Kerr almost coincides with the OUCO of the hairy solution, $r_{\textrm{ISCO}}=7.66$ and $r_{\textrm{OUCO}}=7.44$.

\begin{figure}[htp]
\centering
\includegraphics[width=.4\textwidth]{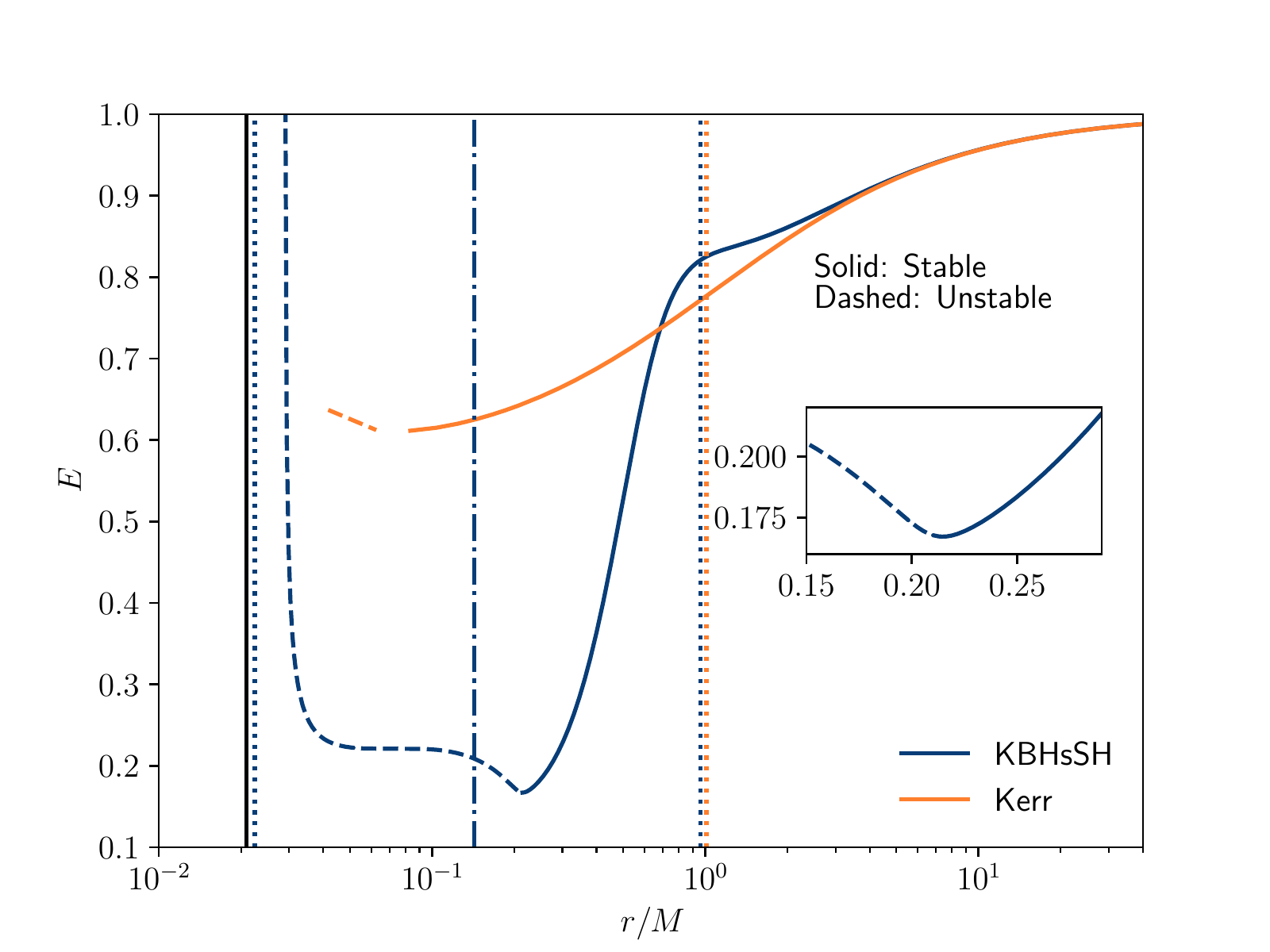}
\includegraphics[width=.4\textwidth]{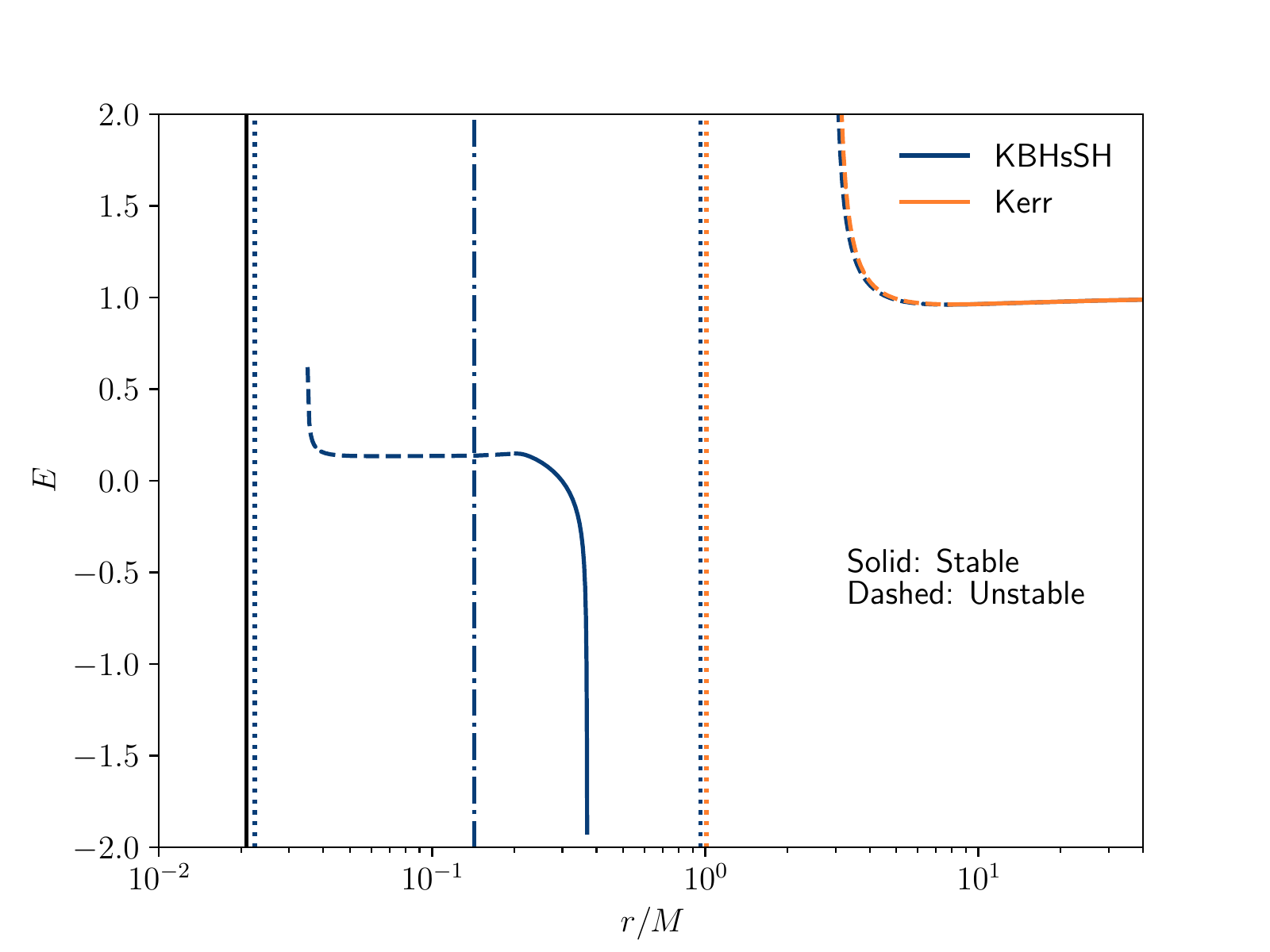}
\caption{Circular orbital energies for Case B. \emph{Left:} $\Omega_+$. \emph{Right:} $\Omega_-$. The vertical lines are the same as in the figure above. Within the exterior ergoregion there are stable circular orbits with negative energy.}
\label{fig:rxErb}
\end{figure}

The energy of the orbiting particle, or compact object is displayed in Fig. \ref{fig:rxErb}. Similarly as before, for the strictly prograde case the instabilities arise once the energy starts increasing as we approach the hole. The $\Omega_-$ eigenvalue case, shown in the right panel as before, unveils that most of the stable orbits within the ergoregion are characterized by negative energy states. In this particular case, at the ISCO the energy is positive but still small, $E=0.15$, as KBHsSH are remarkably efficient in energy conversion. The boost this particle has with respect to a ZAMO at this point is less than $0.12c$, meaning some KBHsSH are also great sources for energy extraction via the Penrose process, in contrast to Kerr BHs that require boosts of over half the speed of light, less likely to happen in astrophysical scenarios.

\subsection{Evolution for A and B}

The circular EMRI evolution within the quadrupole approximation is performed for Cases A and B for both Kerr and KBHsSH in the strictly prograde case with the same initial conditions, where the compact object starts from a distance large enough from the hole so that the differences between both spacetimes in the same case can be negligible. We also evolve the system for retrograde orbits around the hairy black hole of Case A in order to show that the dynamical time can become arbitrarily large, and similarly the signal frequency arbitrarily small. In all the cases, we assume a mass ratio of $\mu=10^{-6}$, and keep all quantities of interest normalized by the BH's mass.

Orbits characterized by the $\Omega_-$ eigenvalue feature an ISCO which is much closer to the hole in the hairy case than for bald Kerr BHs. In the strictly prograde scenario the roles are interchanged, as seen above. The CO reaches much smaller radii around Kerr before plunging in than around a hairy hole. However, the orbital velocity increases monotonically around Kerr, while it can decrease during the approach around a KBHSH, so that at the end of the inspiral it is much less than the one around Kerr. Due to this effect, the inspiralling time can be twice as much in the hairy case.

\begin{figure}[htp]
\centering
\includegraphics[width=.4\textwidth]{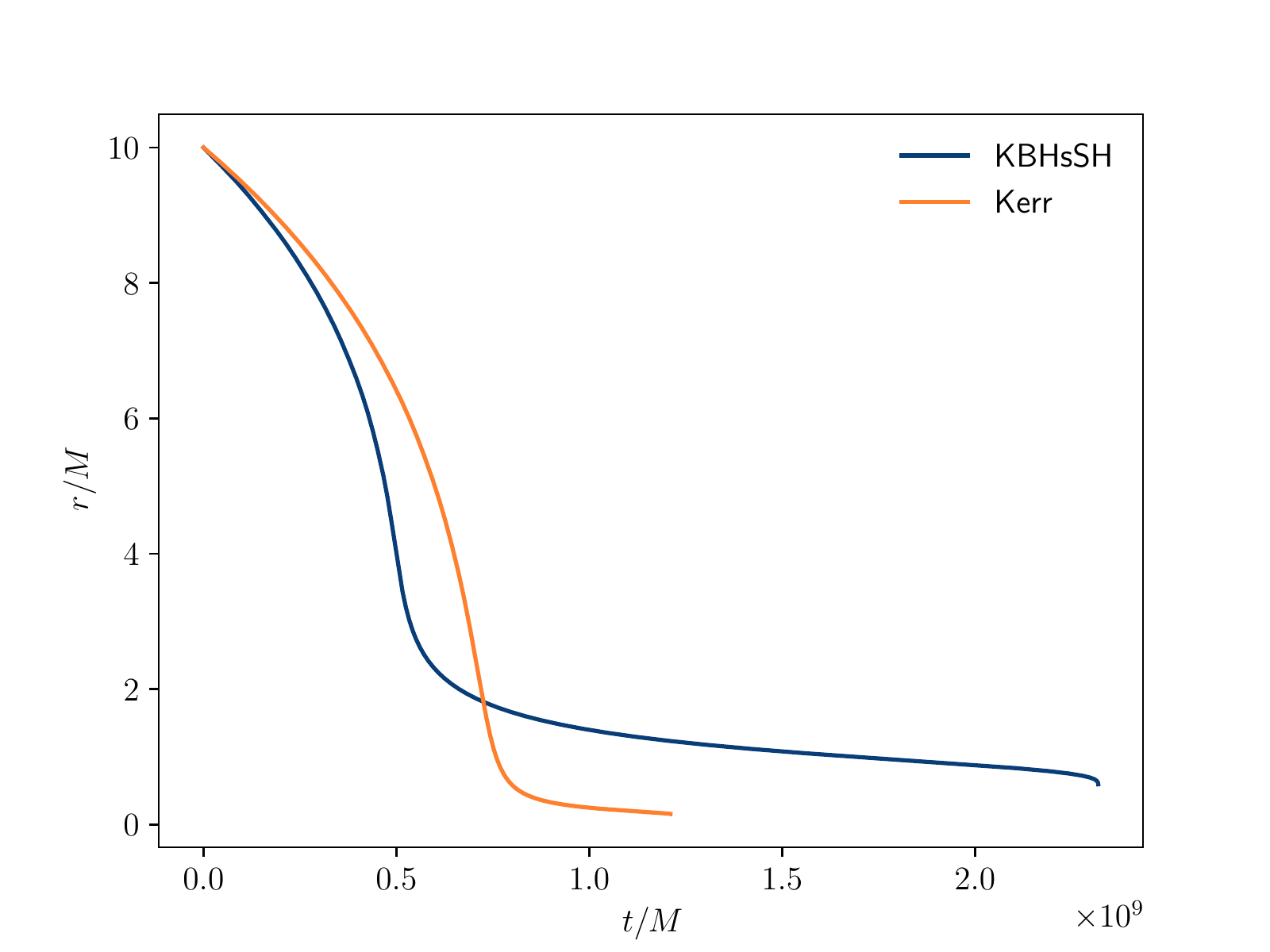}
\includegraphics[width=.4\textwidth]{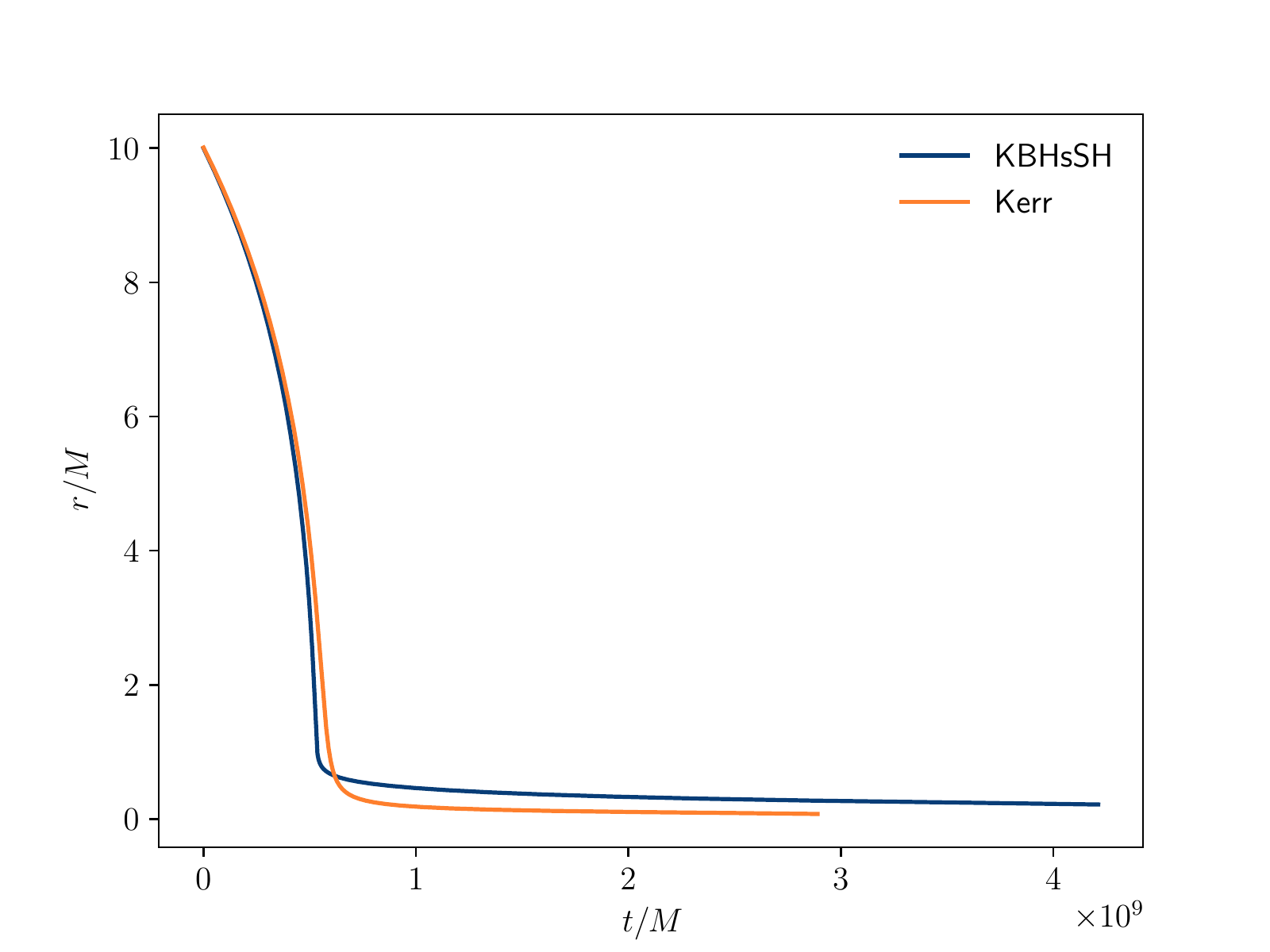}
\caption{Time evolution of the radial coordinate of the prograde inspiraling compact object. \emph{Left:} Case A. \emph{Right:} Case B.}
\label{fig:txr}
\end{figure}

Fig. \ref{fig:txr} shows how $r$ decreases with time for prograde inspirals in Case A (left) and Case B (right), with their respective bald Kerr counterpart. The ISCO for Kerr is much closer to the horizon than it is for hairy BHs in Case A, which the CO reaches in roughly half the time due to the steady increase in the $g_{tt}$ profile. At very late times, we notice an abrupt acceleration increment in the hairy case, right before plunging, which is absent in Case B. This is a consequence of the orbital velocities' behavior and the location of the ISCO, which happens right at the local minimum of $\Omega_+$ in case B, but right after the velocity starts increasing again in case A.

\begin{figure}[htp]
\centering
\includegraphics[width=.4\textwidth]{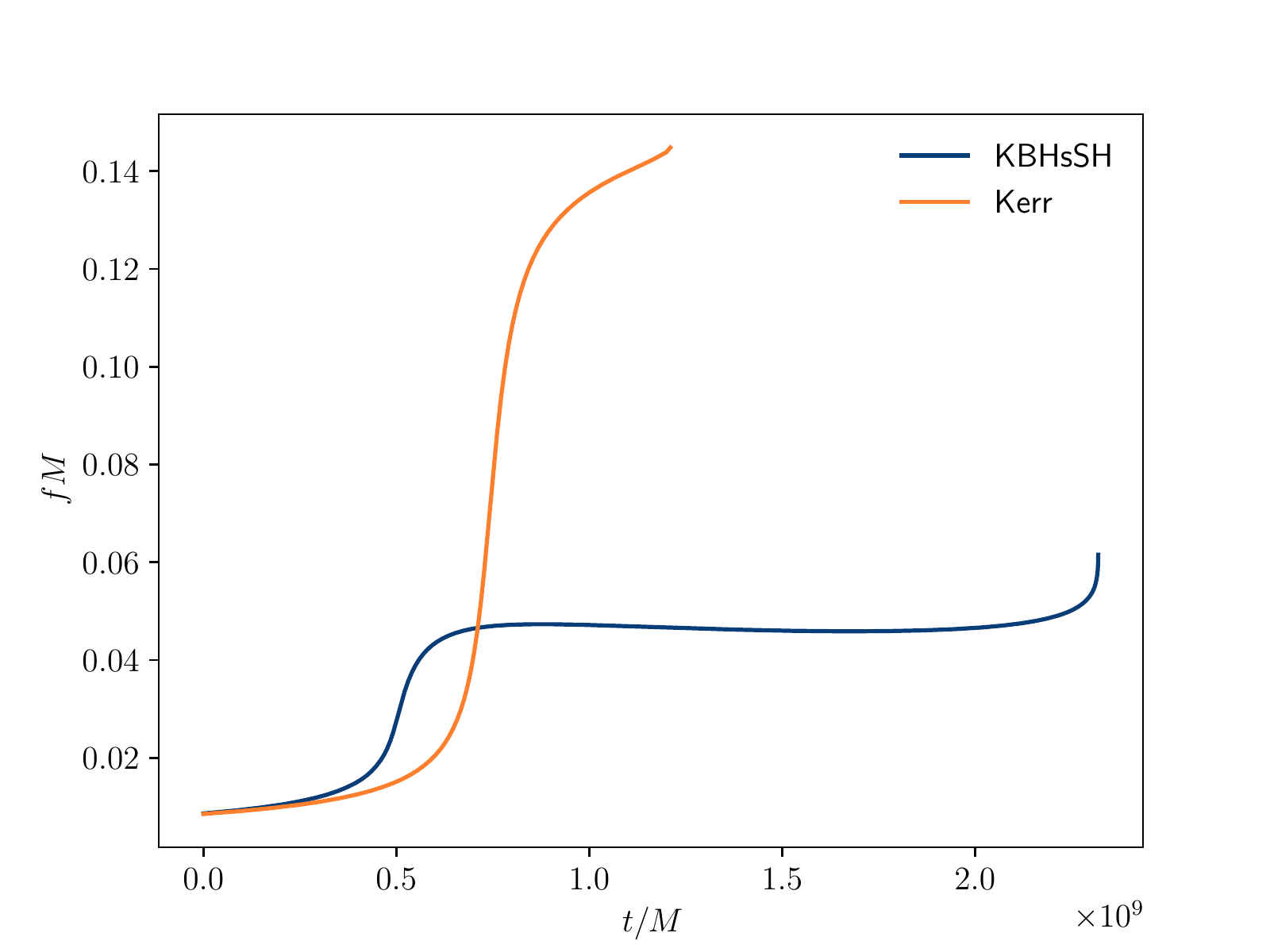}
\includegraphics[width=.4\textwidth]{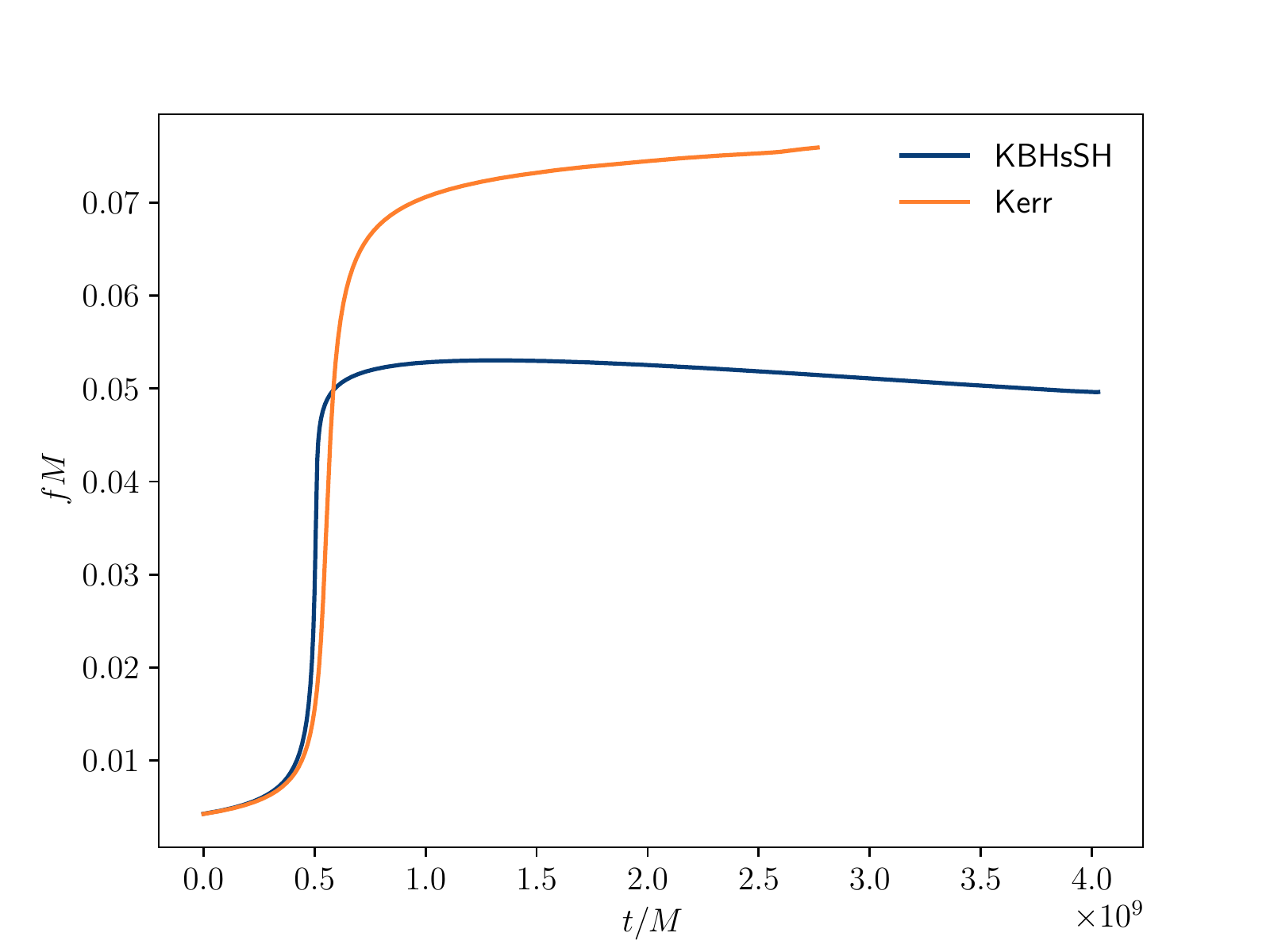}
\caption{Time evolution of the GW frequency of the prograde inspiraling compact object. \emph{Left:} Case A. \emph{Right:} Case~B.}
\label{fig:txf}
\end{figure}

The observable frequency evolutions throughout the inspirals are depicted in Fig. \ref{fig:txf}. As we anticipated from the profile of the orbital velocities, these KBHsSH feature a backward chirping as the CO moves inwards through the scalar hair. Because these solutions are quite \emph{hairy}, most of the mass and angular momentum are stored in hair rather than in hole and the frequencies slightly drop once the CO leaves behind the region of highest energy density. For similar reasons, they grow fairly smaller than those of CO orbiting Kerr BHs of same $(r_H, M)$ parameters. During the last part of the evolution before plunging, we note a transition to forward chirping in case A which is absent in case B.

\begin{figure}[htp]
\centering
\includegraphics[width=.4\textwidth]{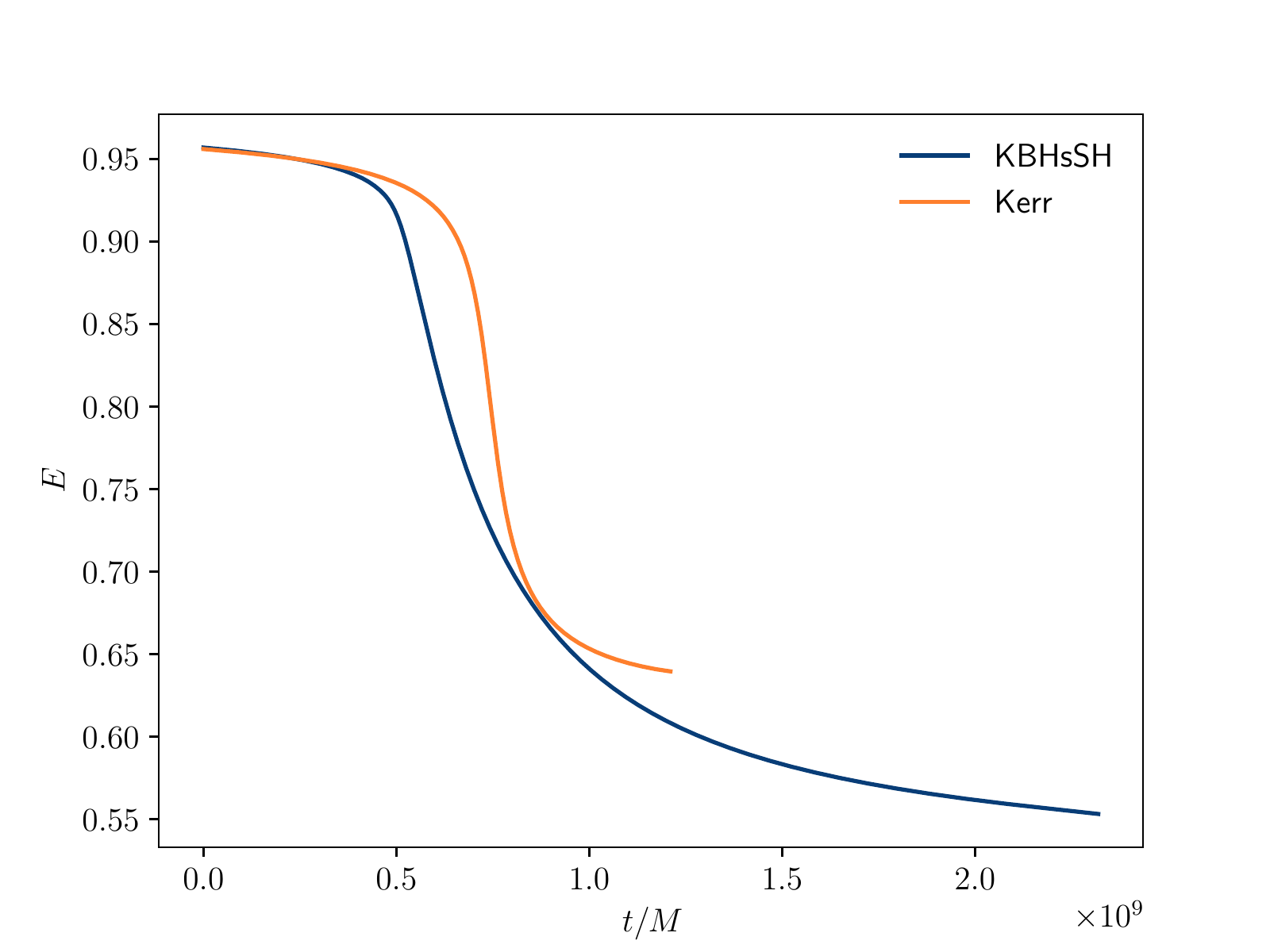}
\includegraphics[width=.4\textwidth]{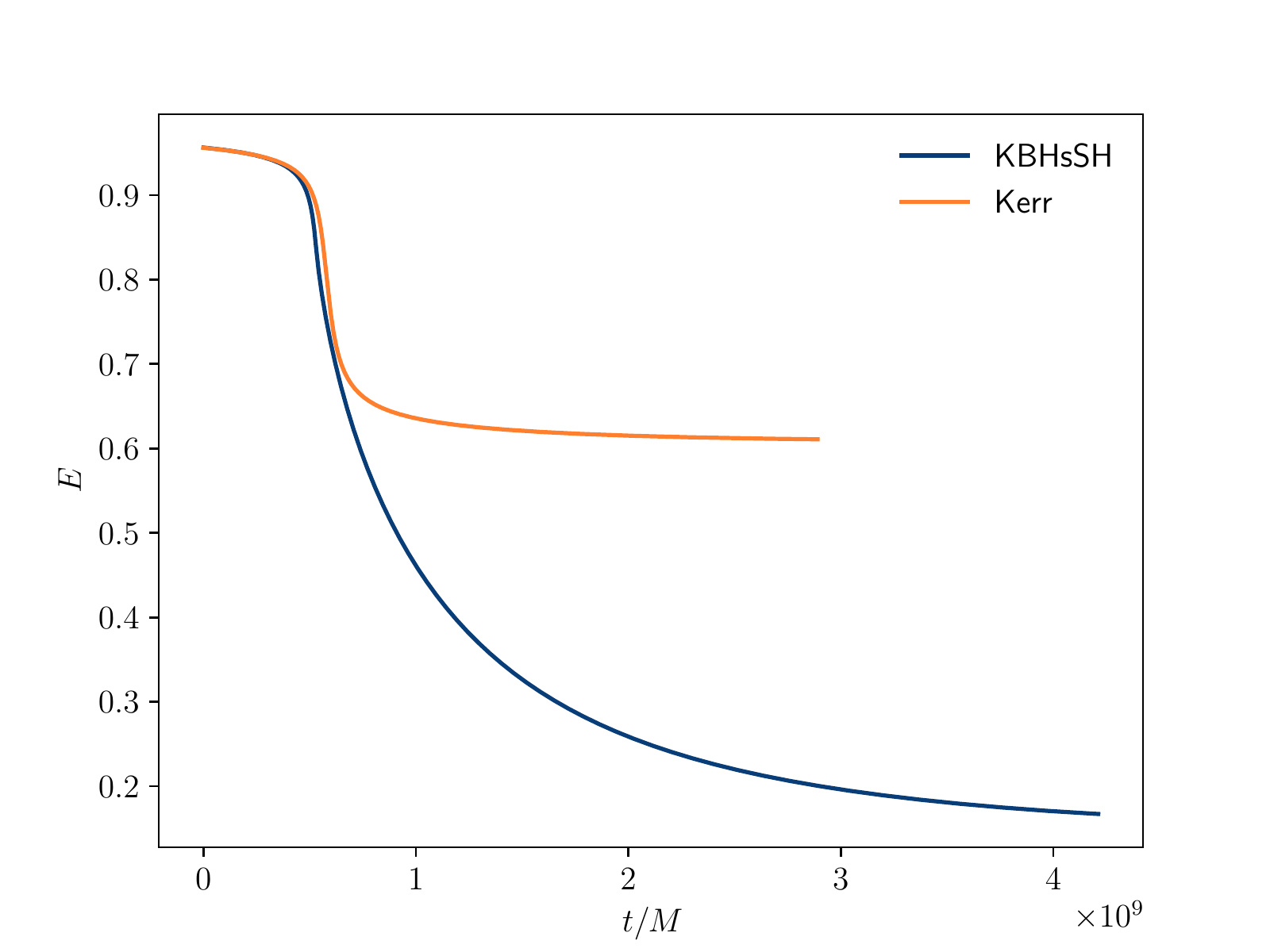}
\caption{Time evolution of the energy of the prograde inspiraling compact object. \emph{Left:} Case A. \emph{Right:} Case B.}
\label{fig:txe}
\end{figure}

\begin{figure}[htp]
\centering
\includegraphics[width=.4\textwidth]{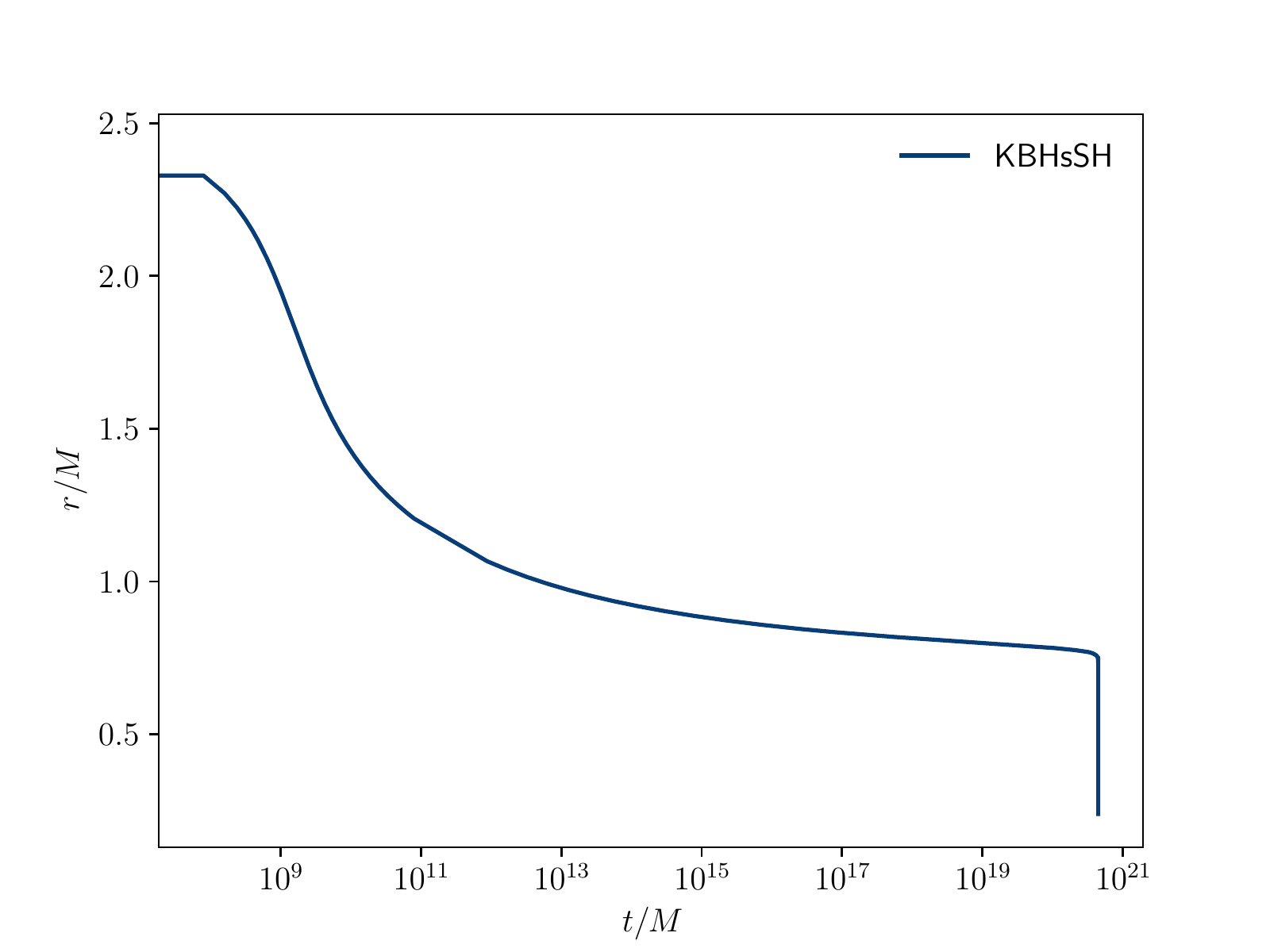}
\includegraphics[width=.4\textwidth]{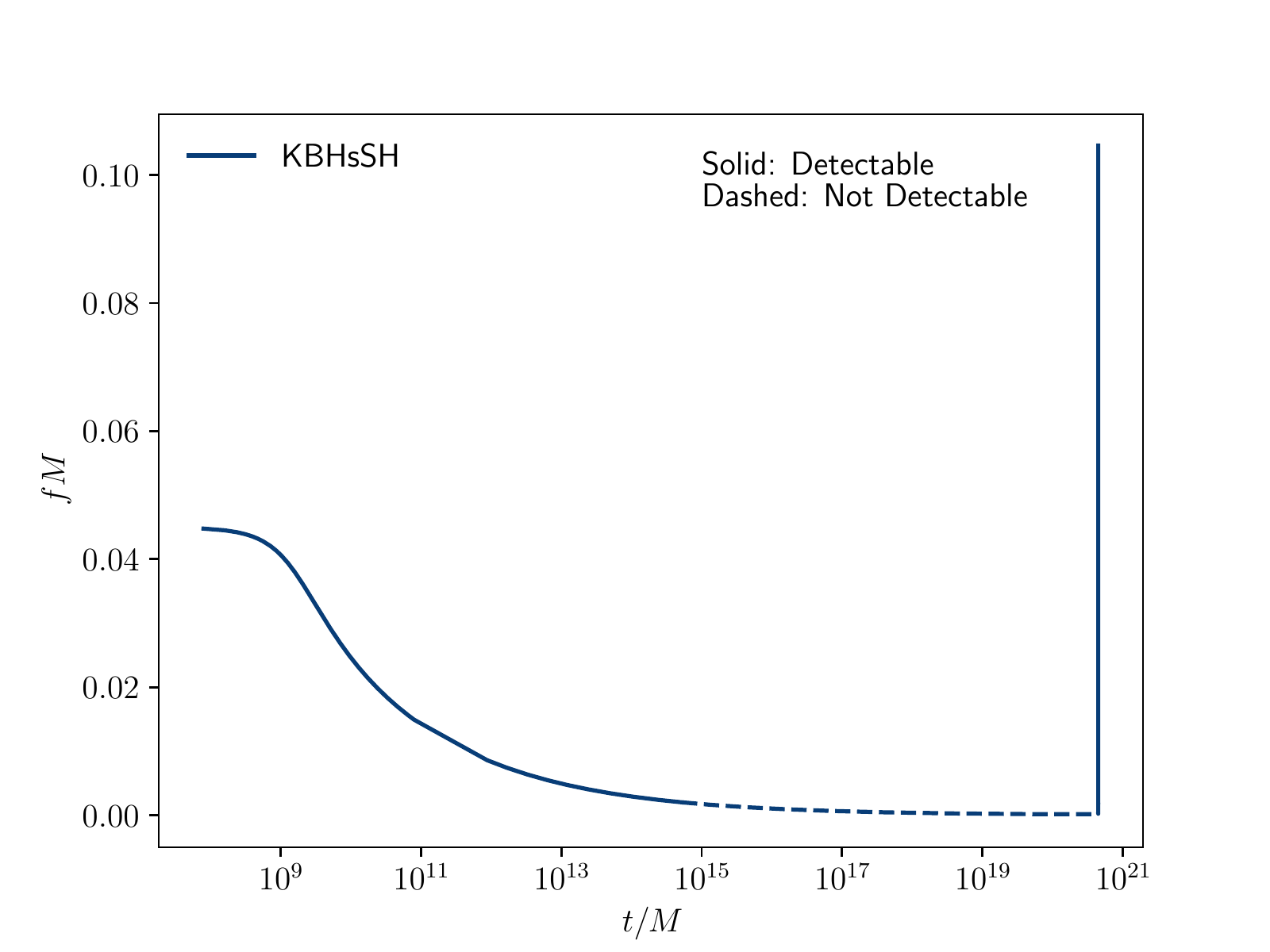}
\caption{Retrograde evolution for Case A. \emph{Left:} Orbital radius against time. \emph{Right:} GW frequency against time. Assuming the central object to be as massive as Sagittarius A*, the dashed interval of the curve falls below LISA's sensitivity range. The whole evolution shown happens over a timespan roughly ten times the age of the Universe for a central object whose mass is comparable to that of Sagittarius A* and this is due to the fact that the BH in Case A is taken to be very close to developing a SR. }
\label{fig:txrf}
\end{figure}

The retrograde evolution in Case A is displayed in Fig. \ref{fig:txrf} for illustration purposes only.  We emphasize that the solutions deform continuously through the set, i.e. connecting cases A and C, and therefore one might find KBHsSH arbitrarily close to forming a SR. Once this happens, taking only the quadrupole mode into account as in this approximation of circular orbits inspirals, the evolution stops entirely as the fluxes disappear when $\Omega_-=0$. Hence, a CO initially at a radius $r>r_{st}$ will evolve until the SR, with decreasing orbital velocity until it reaches this point and stops, never to plunge into the hole. In this particular example case, the plunging occurs, but as the CO approaches the region of almost zero slope in the $g_{tt}$ component, the orbital velocity becomes extremely small, reflecting the backwards chirping we see in the plotted frequency, which falls out of the observable range of LISA. Once this region is finally overcome, the evolution assumes a more intuitive behavior, and the frequency grows until the plunging. If the central object has a mass comparable to Sagittarius A*, the evolution we considered (from $r_0\sim 2.4M$) would take roughly ten times the age of the Universe. This, of course, highly dependents  on how close to the formation of a SR the hairy black hole is since at a SR the evolution practically stops as explained above. Furthermore, for a good part of the evolution, the signal falls below LISA's threshold frequency of detectability of $0.1$mHz as indicated by the dashed pattern in the figure, with a minimum frequency of $7.2\mu$Hz achieved at the maximum of $g_{tt}$. All other hairy cases here presented have signal frequencies of the same order of magnitude as their Kerr counterpart, and therefore fall within LISA's range for the same range of BH's masses.

\subsection{Case C}

\begin{figure}[htp]
\centering
\includegraphics[width=.4\textwidth]{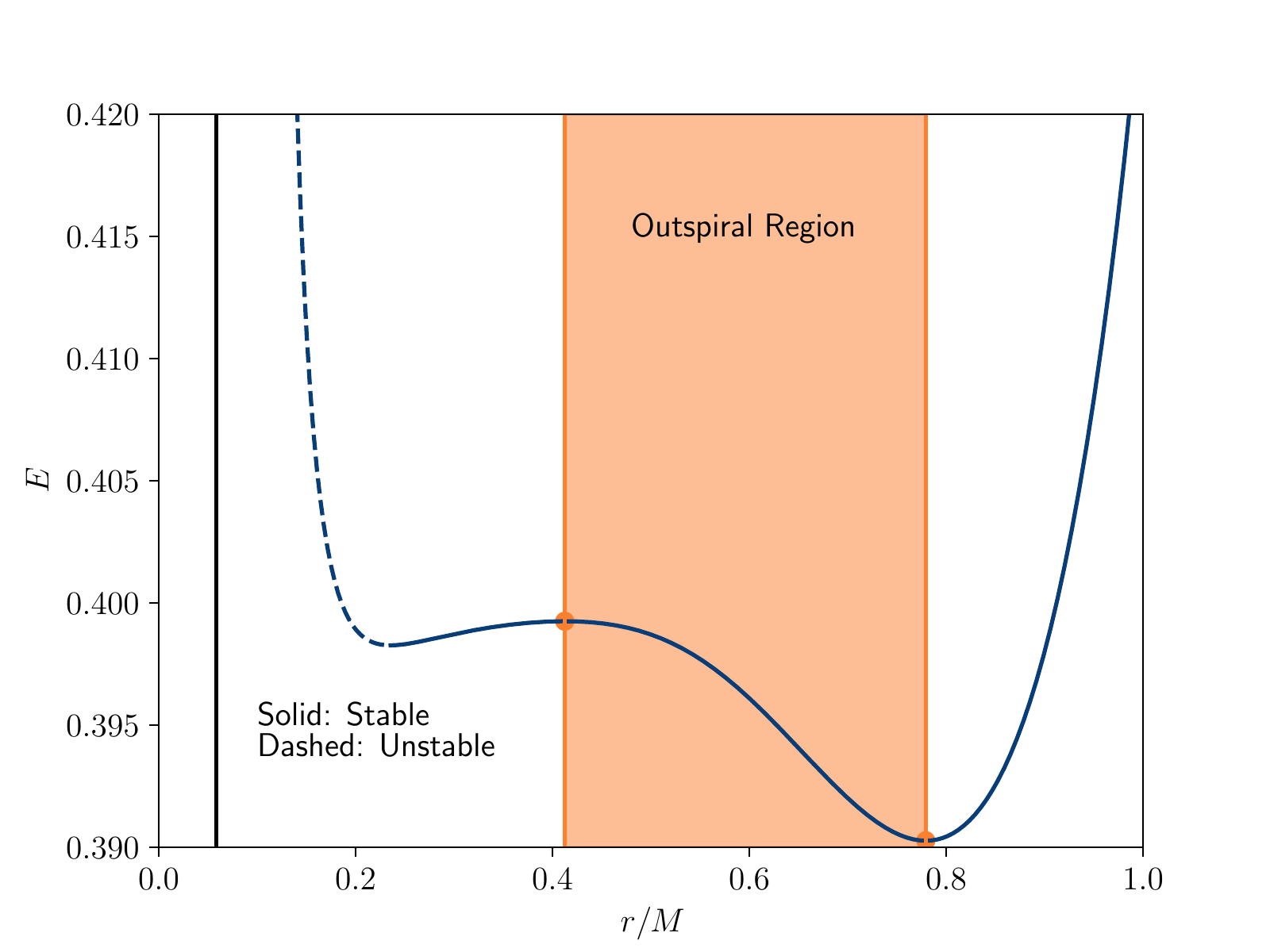}
\caption{Energy of the CO against the radial coordinate in the presence of two SRs. Three local extreme are found: the innermost occurs at the ISCO, and the following to at the SRs. Within the last two the energy has a negative slope and therefore a radiating CO in circular orbit in this region would outspiral until reaching the outermost ring where the fluxes stop.}
\label{fig:rxEst}
\end{figure}

The third solution we consider features two SRs, and therefore a CO approaching from beyond the outermost one would stop at it as described above. Performing an evolution scheme would provide no more information than can be inferred from the orbital parameters. Nevertheless, we would like to draw attention to another exclusive feature of spacetimes containing these rings, namely a possible outspiral. In Fig. \ref{fig:rxEst} we show the circular orbiting CO's energy against the radial coordinate. Within the shown region there are three local extreme for the energy. The innermost coincides with the ISCO and hence a change in orbital stability. The other two (maximum in the middle and minimum for the outermost) correspond to the SRs. Remarkably, for this particular solution those are both withing the stable interval. If a CO in a initially eccentric orbit manages to reach the region within the rings and circularizes there, radiating energy via the fluxes would cause it to move away from the hole, outspiriling until it reaches the outermost ring, where the fluxes stop.

The existence of both in- and outspirals ensures that the spacetime also hosts \emph{floating orbits} from different flavors. In the present work, we assumed that all emissions come solely from the CO. In such case, there is a floating orbit at each static ring. If one is to consider background perturbations, the orbiting CO would induce a loss in the black hole mass that could equal its own emission, as first pointed out by Misner \cite{PhysRevLett.28.994}, creating a balanced system in which both radiation effects cancel out and the orbit does not shrink. For KBHsSH, the hole's flux would be of order
\be
\label{eq:dmdtfo}
\dot{E}_H\sim\frac{\left(r^2\Omega\right)^5}{\omega_s-\Omega},
\ee
which actually overcomes the quadrupole flux from the CO in certain regions of the spacetime of several solutions. In particular as $\Omega$ decreases in absolute value with decreasing $r$. A much more meticulous analysis is required in order to weight out the contributions of the influxes through the event horizon. Moreover, one needs to consider scalar perturbations due to the hair, which should cause further surperradiant phenomenon, as discussed in detail in \cite{PhysRevLett.107.241101}. Note, however, that for KBHsSH the scalar field couples only minimally to gravity.

\begin{figure}[htp]
\centering
\includegraphics[width=.4\textwidth]{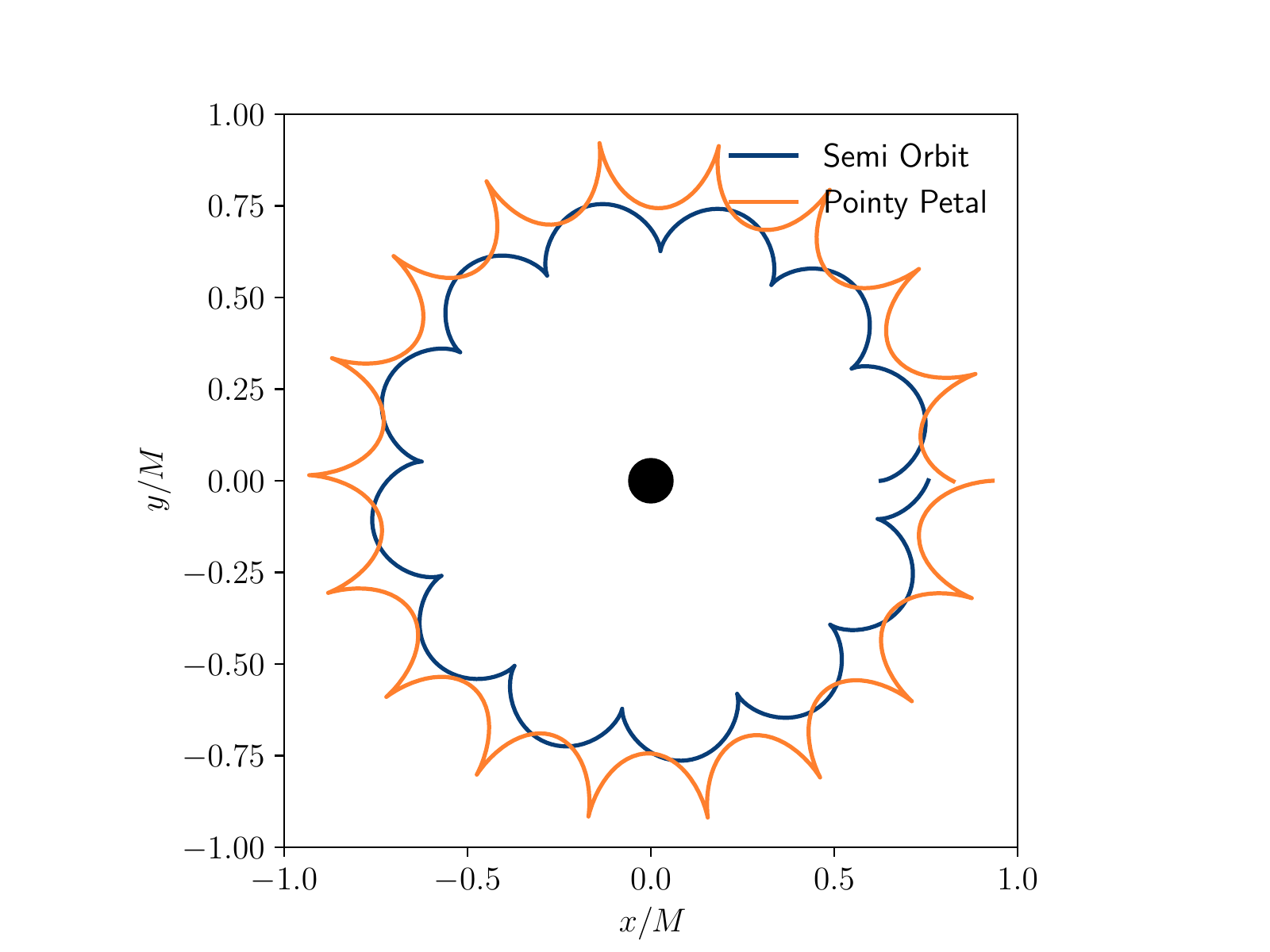}
\caption{Two peculiar types of orbits on the equatorial plane, typical of spacetimes containing a static ring. The pointy-petal orbit is retrograde, and the CO is periodically momentarily at rest at a radius larger than $r_{st}$. The semi orbit is prograde and the CO is periodically instantaneously at rest at a radius smaller than $r_{st}$.}
\label{fig:semi_pointy_orbits}
\end{figure}

A peculiar class of orbits found in spacetimes warped by rotating BSs \cite{Collodel:2017end,PhysRevD.90.024068,Grould_2017} is also present in KBHsSH containing SRs. They are characterized by orbits where the particle (or CO in our case) periodically finds itself momentarily at rest and come in two kinds, discerned by the location where the rest points occur. Orbits for which the particle is at rest at the perastrons are called \emph{semi orbits}, while those for which it comes to a halt at the apoastrons are called \emph{pointy-petal orbits}, due to the curve they draw in a parametrized two dimensional space, as disclosed in Fig. \ref{fig:semi_pointy_orbits}. The SR dwells between the rest points of each orbit, and the pointy-petal is in retrograde motion while the semi orbit in prograde (the sing of the eigenvalue $\Omega_-$ flips upon crossing the SR). No analogue class of orbits is found in pure KBH spacetimes.

\begin{figure}[htp]
\centering
\includegraphics[width=.4\textwidth]{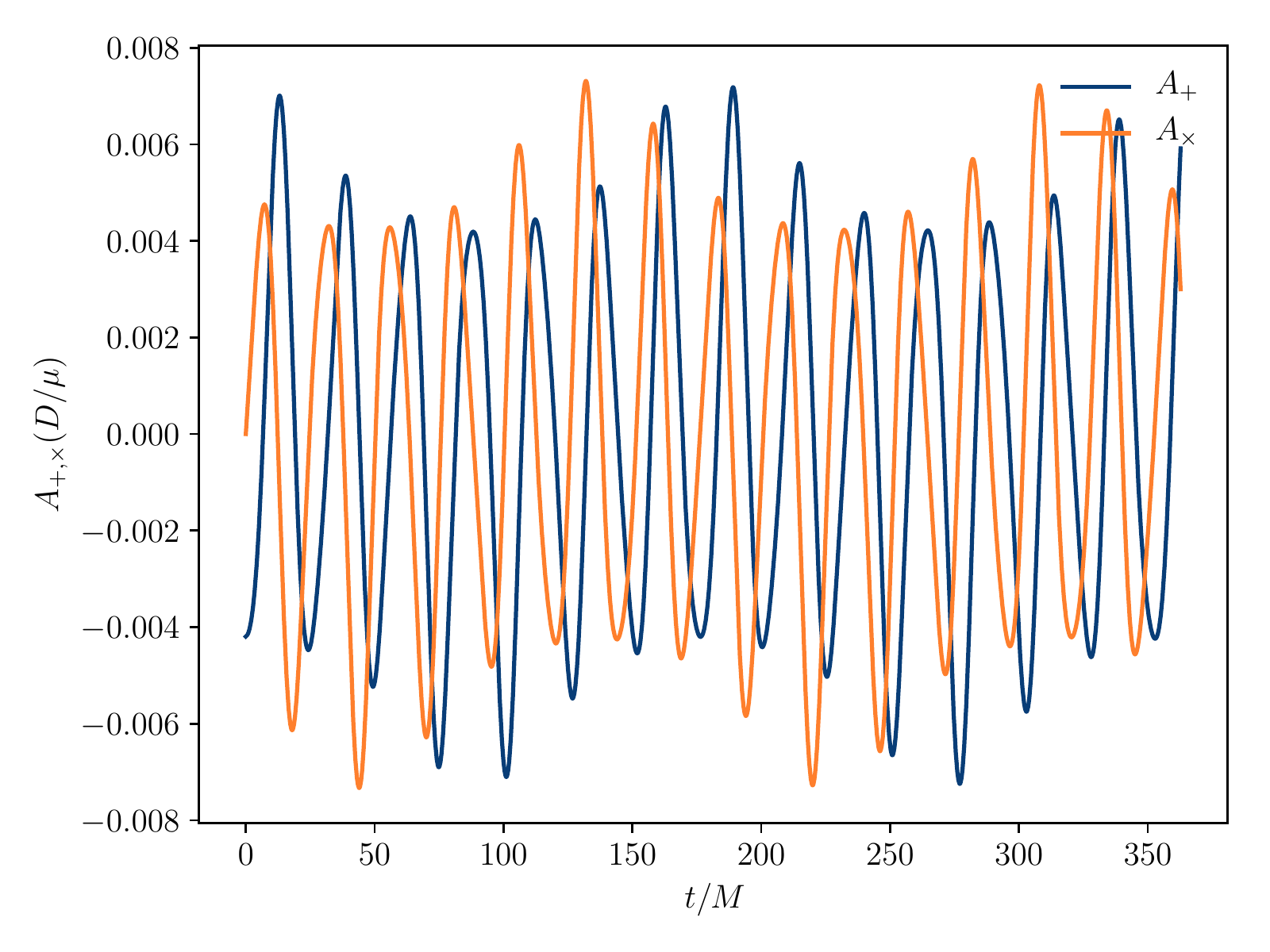}
\includegraphics[width=.4\textwidth]{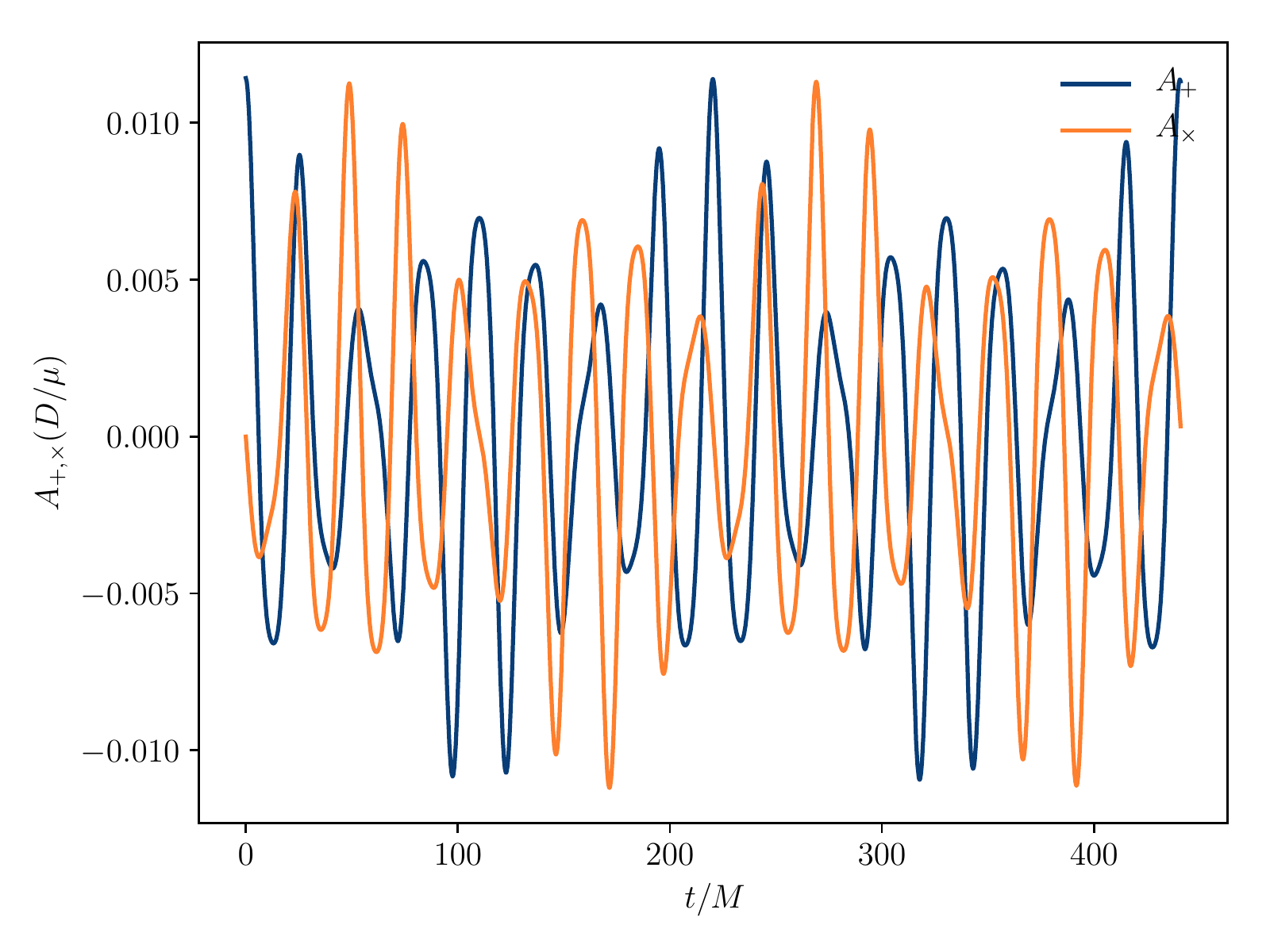}
\caption{Amplitudes calculated with the quadrupole approximation. \emph{Left:} Semi orbit. \emph{Right:} Pointy-petal orbits. }
\label{fig:a+a-}
\end{figure}

\begin{figure}[htp]
\centering
\includegraphics[width=.4\textwidth]{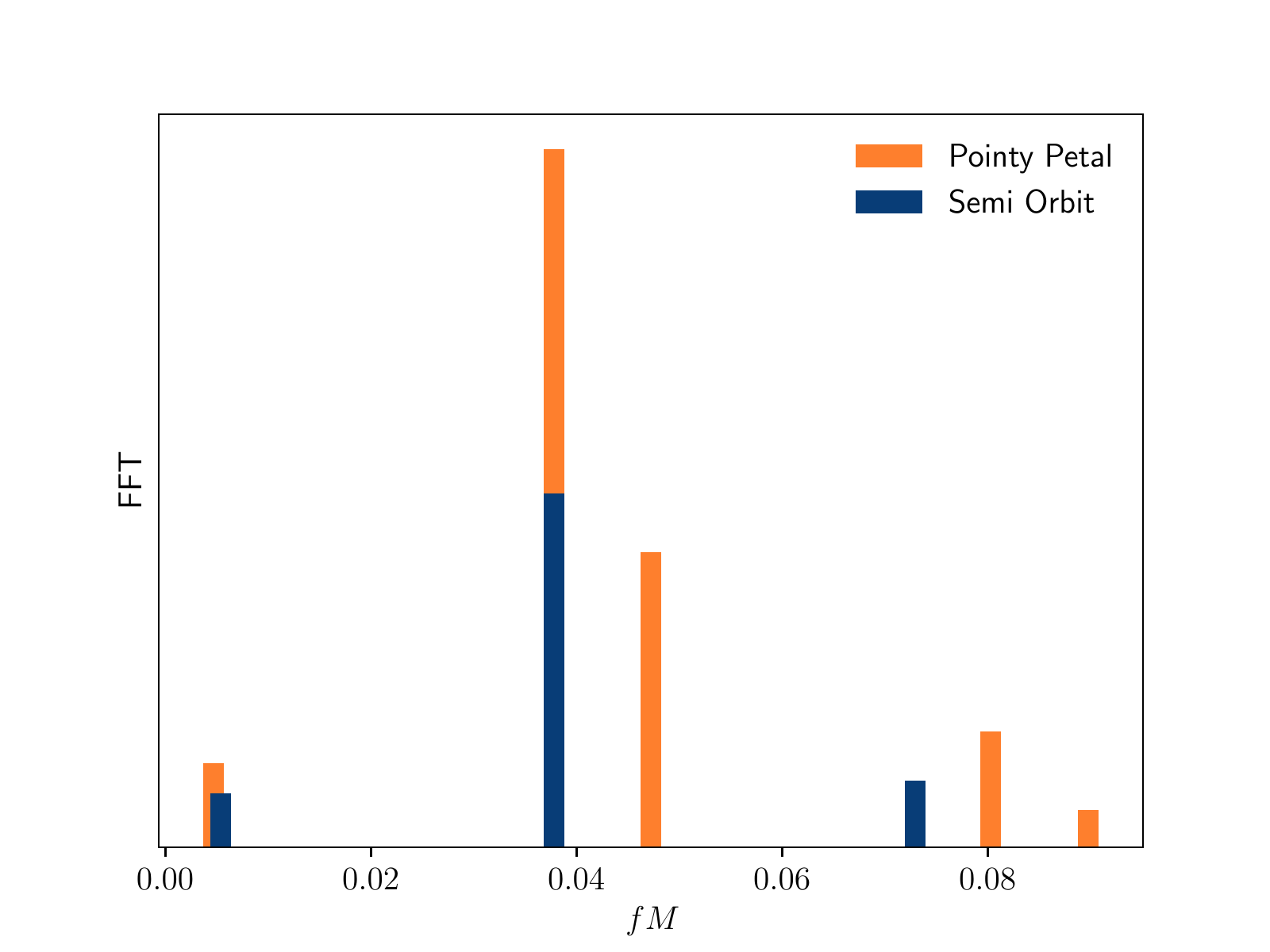}
\caption{Fast Fourier Transform (FFT) of the GW signals of the pointy-petal and semi orbit.}
\label{fig:fft}
\end{figure}

It is fairly reasonable to expect that a CO following such exquisite orbits would leave imprints on its emitted GW characteristic of the rest points, which could be used to probe the existence of these BHs. Below, in Fig. \ref{fig:a+a-} we present the waveforms for each orbit over a whole cycle around the central object by using eqs. (\ref{eq:a+}) and (\ref{eq:ax}), but neglecting the fluxes due to the narrowness of the timespan. As these are not meant to be used as real templates for future observations due to the limitation of this approach, we do not perform any quantitative analysis on the outcome, but rather entertain what strong features it unveils that would in principle be present qualitatively also in the real signal. In both cases, the signal from the second half revolution ($\varphi\in[\pi,2\pi]$) is identical to the first ($\varphi\in[0,\pi]$), as expected from the location of the observer. Furthermore, splitting the signal of each half revolution into two, the second quarter mirrors the first one. The signals differ considerably between each type of orbit, and in the case of the pointy-petal one, we notice a localized change of frequency that results in a bulgy signal in each quarter revolution. As a consequence, the spectrum of the pointy-petal is richer than the semi orbit, as shown in Fig. \ref{fig:fft} where we display the frequencies for both signals.

\section{Conclusions}
\label{sec:conclusions} 

In this work we have identified interesting inspirilling and GWs properties of equatorial plane EMRIs in KBHsSH spacetimes within the hybrid formalism, considering only quadrupole moments contributions to the fluxes. By selecting particular solutions which share major features of spacetimes warped by spinning solitons and BHs, we were able to highlight specific traits of circular and noncircular orbits which are absent in KBHs and that produce imprints on the GW signal of an inspirilling CO.

In essence, when enough energy is stored in the solitonic hair, its off-center distribution causes the metric functions to portrait a nonmonotonic profile on the equatorial plane. In turn, the circular orbit angular velocity also behaves nonmonotonically. This creates a backwards chirping in the GW signal frequency emitted by the CO, which does not appear for CO circularly inspirilling towards a KBH. When comparing KBHsSH with KBHs of same mass and horizon radius, the signals emitted by a CO around Kerr present overall larger frequencies than around KBHsSH, but those can be considerably smaller for the early stages of the evolution.

If a static ring is present, within the quadrupole approximation, the fluxes become trivial at that point for a CO in circular orbit with $\Omega_-$ orbital velocity, and the evolution stops, i.e. the CO remains at rest at this ring not emitting any GW. On the other hand, solutions that get arbitrarily close to forming a static ring serve as stage for COs that can orbit the central object with $\Omega_-$ arbitrarily close to zero and the evolution timespan can be several times larger than the age of the Universe for a KBHSH with galatic center proportions. In such case, part (but no all) of the emitted signal can fall out of the LISA frequency range, as the absolute value of $\Omega_-$ becomes too small in an intermediate region of the inspiral. As for the energy of the CO in circular orbit, it can feature multiple local extrema. For example, for one of the selected solutions we consider (Case C) two static rings are present and the energy profile has a local maximum at the inner ring and a local minimum at the outer one. If a sequence of events brings a CO to circularize with orbital velocity $\Omega_-$ in the region bounded by these extrema, it would outspiral towards the outer ring where once more the fluxes stop. 

A particular class of equatorial noncircular orbits were also analyzed, namely pointy-petal and semi orbits. These are only realized in spacetimes containing static rings and are characterized by the CO being periodically momentarily at rest with respect to the fiducial observer. Without considering radiation reaction, we constructed waveforms for a whole revolution considering both orbits, again with quadrupole approximation. In special for the pointy-petal case, we note that periodic protuberances appear in the signal, providing a richer spectrum of frequencies.

The current paper makes the first step in the investigation on EMRIs in KBHsSH spacetimes. To our knowledge, there is yet no work in the literature investigating EMRIs in rotating spacetimes whose solutions are only known numerically. The main purpose of this research was to identify particular solutions of interest and pinpoint what peculiarities they might reserve which could provide significant imprints in the GW signals not present in the Kerr case. In a future work, we intend to go further by using more sophisticated and general approaches that will allow us to check up to what extent the qualitative and quantitative obeservations in the present paper remain true. In particular, we will be interested whether eccentricities could be developed quenching the stop of the fluxes and possibly the outspiral region. In another framework, by adopting the generalized Teukolsky formalism, accurate waveforms could be generated and the existence of floating orbits could also be reliably probed.

\section*{Acknowledgements}
LC and DD acknowledge financial support via an Emmy Noether Research Group funded by the German Research Foundation (DFG) under grant no. DO 1771/1-1. SY would like to thank the University of Tuebingen for the financial support.  SY acknowledges financial support by the Bulgarian NSF Grant KP-06-H28/7. Networking support by the COST Actions  CA16104 and CA16214 is also gratefully acknowledged.
 
\bibliographystyle{ieeetr}
\bibliography{biblio}
\end{document}